# DNA Bending and "Structural" Waters in Major and Minor Grooves of A-tracts. Monte Carlo Computer Simulations

Alexander V. Teplukhin[*], Valery I. Poltev[‡] and Victor B. Zhurkin[§]

[*] Institute of Mathematical Problems of Biology, Russian Academy of Sciences, Pushchino, Moscow Region, 142292, Russia

[‡] Institute of Theoretical and Experimental Biophysics, Russian Academy of Sciences, Pushchino, Moscow Region, 142292, Russia

[§] Laboratory of Mathematical Biology, NCI, NIH, Bg. 12B, Rm. B116, MSC 5677, 12 South Dr., Bethesda, MD 20892-5677

E-mail: tepl@impb.psn.ru

ABSTRACT

To elucidate the possible role of structural waters in stabilizing bent DNA, various conformations of AT-containing decamers, $(A_5T_5)_2$ and $A_{10}:T_{10}$, were studied by Monte Carlo simulations. The duplexes were constrained to reproduce the NMR inter-proton distances for the A-tracts at two temperatures: ~5° and ~35°C. Analysis of the water shell structures revealed a strong correlation between the groove widths on the one hand, and the types of hydration patterns and their probabilities on the other hand. Depending on the minor groove width, the following patterns were observed in this groove: either an interstrand "hydration spine", or interstrand two-water bridges, or a double ribbon of intrastrand sugar-base "water strings". Hydration shell in the major groove is less regular than in the minor groove, which agrees with crystallographic data. As in previous studies, energetically advantageous hydration is found for the A-tract conformations with narrow minor groove and wide major groove (B'-like DNA), known to be correlated with DNA bending and curvature. In addition, our calculations indicate that the advantage of such DNA conformations is coupled with the increase in adenine N7 hydration in the major groove. As the major groove is opened wider, its hydration shell is enriched by energetically favorable "trident waters" hydrogen bonded to three hydrophilic centers belonging to two adjacent adenines: two N7 atoms and one amino-proton H(N6). Based on these results, we suggest that formation of the novel hydration pattern in the major groove is one of the factors responsible for stabilization of the B'-like conformations of A-tracts at low temperature in the absense of dehydrating agents, leading in turn to the strong intrinsic curvature of DNA containing alternating A-tracts and "mixed" GC-rich sequences. This concept is consistent with the available experimental data, and is furthermore testable by $^{15}N$ NMR.

# INTRODUCTION

DNA, as a linear molecule, is quite susceptible to the environment, each nucleotide being substantially exposed to solvent. At the same time, the internal interactions stabilizing any one particular conformation of DNA, are relatively weak, compared to globular proteins. As a consequence, DNA conformations are aptly responsive to external effects, and the local structure of DNA can vary over a wide range. In turn, the structural variability of DNA, such as the sequence-specific bending and curvature, is important for a number of basic processes, including protein binding and gene regulation (for reviews see Trifonov, 1986; Crothers et al., 1990; Hagerman, 1990, 1992; Harrington and Winicow, 1994; Travers, 1995; Olson and Zhurkin, 1996).

Bent DNA is characterized by periodically modulated sizes of the major and the minor grooves (Drew and Travers, 1985; Burkhoff and Tullius, 1987). The atom arrangements in the grooves differ substantially for different sequences; thus, the possibility exists that some of sequences can be solvated with more structural waters than others. This raises the important question of the correlation between the sequence-dependent structures and energetics of DNA duplex on the one hand, and the features of its hydration shell on the other hand. In particular, it is of interest to examine the role of solvation effects in stabilizing the bent conformations of the double helix.

To address this question, we have studied structural water patterns formed in the DNA grooves in a sequence-specific fashion, and compared them to bulk waters. We have chosen Monte Carlo (MC) simulations as a tool to reveal the most favorable modes of solvation of DNA, and have analyzed duplexes containing the so-called A-tracts (contiguous runs of several adenines in one strand). The reason for doing so is given below.

Most of the curved sequences studied so far, starting historically with kinetoplast DNA, contain A-tracts periodically alternating with "mixed" GC-rich spacers (Marini et al., 1982; Diekmann, 1986; Koo et al., 1986; Hagerman, 1988; Beutel and Gold, 1992). Strong DNA curvature is likely to be a consequence of the difference in conformation and flexibility between the A-tracts and the GC-rich spacers. Among other factors, solvation effects could be related to this difference. Currently there is no consensus in the literature regarding explanation of this phenomenon at the atomic level. Here we mention only briefly the well known discrepancies between the "A-tract" or "junction" model (Crothers et al., 1990; Haran et al., 1994), the "non-A-tract" model (Calladine et al., 1988; Goodsell et al., 1994; Bruckner et al., 1994) and various "wedge" models, both static and flexible (Bolshoy et al., 1991; De Santis et al., 1990; Zhurkin et al., 1991; Olson et al., 1993). All these models, however, have one feature in common, which appears to be consistent with the whole multitude of experimental data, namely, that the A-tracts in solution are characterized by a more negative roll of base-pairs and compressed minor groove, compared to the "mixed" GC-rich sequences.

At the macroscopic level of hundreds of base pairs, the DNA curvature is easily detectable by retardation on polyacrylamide gel. This retardation crucially depends on the environment: type of cations, activity of water and temperature. Divalent cations increase DNA retardation on gels (Marini et al., 1984; Shlyakhtenko et al., 1990; Nickol and Rau, 1992; Brukner et al., 1994). By contrast, addition of alcohol (EtOH and MPD) and increase in temperature gradually diminish the curvature of the kinetoplast DNA (Marini et al., 1984; Sprous et al., 1995). Notice, however, that when the "spacer" sequences located between the A-tracts are AT-rich ($T_5$, TATAT or TCTCT), the temperature dependence is more complicated, and the maximum curvature is observed at 25°-35°C (Diekmann, 1987; Shlyakhtenko et al., 1990, 1992). All these data, taken together, reveal the role of solvent in moderating macroscopic DNA bending and curvature.

At the local level, the NMR data indicate that the conformation and dynamics of the A-tracts are different at low temperature (5°-15°C) and at room temperature or higher (Leroy et al., 1988; Nadeau and Crothers, 1989; Katahira et al., 1990). Traditional interpretation attributes this difference to the changes in the base-pair inclination (Marini et al., 1984), in particular, to a negative roll at low temperature (Lipanov and Chuprina, 1987; Nadeau and Crothers, 1989). This interpretation is consistent with the temperature dependence of the NMR and gel electrophoresis measurements mentioned above. By analogy with the X-ray fiber structure of *poly(dA):poly(dT)* also having negative roll of bases and a narrow minor groove, the low temperature conformation of the A-tract in solution is denoted B' form (Arnott et al., 1983; Alexeev et al., 1987; Chuprina, 1987).

Several detailed hypotheses have been developed to elucidate the unusual properties of $A_n:T_n$ tracts in solution, but none of them accounts for all available data (Diekmann et al., 1992). One of them, the "hydration spine" hypothesis, considers a regular positioning of the "structural" waters in the minor groove as the main factor stabilizing the B'-like conformation of A-tracts in solution (Chuprina, 1985, 1987). This model, based on the crystal structure of the B-DNA dodecamer (Drew and Dickerson (1981), successfully predicted the larger number of bound waters in $A_n:T_n$ tracts compared to the alternating A-T sequences (Buckin et al., 1989). It remains unclear, however, *why* the regular hydration spine in the narrow minor groove should be energetically preferable compared to water strings in the relatively wide minor groove of the "standard" B-form (Prive et al., 1987).

In other words, it is not yet understood what is "the primary determinant of the minor groove width": whether formation of the hydration spine leads to narrowing of the groove (Chuprina, 1987) or, on the contrary, the minor groove is narrowed as a result of other interactions, such as base stacking (Diekmann et al., 1992), with the water spine "forming as a consequence" (cited from Hagerman, 1992).

Furthermore, this concept based exclusively on the formation of the hydration spine in the minor groove, fails to explain why the DNA curvature depends upon the nature of the exocyclic groups in the purine 6- and 7-positions in the major groove (Diekmann et al., 1987; Seela et al., 1989). The hydration of the minor groove alone cannot account for the noticeable difference in the electrophoretic nobilities of the DNAs containing $A_4{:}T_4$ and $(AATT)_2$ tracts (Koo et al., 1986; Hagerman, 1988). In addition, the amount of water released upon binding of netropsin to *poly(dA):poly(dT)* is substantially larger than would be expected if the frozen-in water melted only from the minor groove (Marky and Kupke, 1989). The optical, spectroscopic and calorimetric studies of the B' → B "premelting" transition in the $A_n{:}T_n$ fragments also point to melting and release of "bound" water into the bulk solvent (Herrera and Chaires, 1989; Chan et al., 1990). So, a thorough examination of DNA hydration, depending on DNA conformation, especially in the major groove, is necessary. One more reason to pay attention to the major groove is that the third strand of the triplex is located there (Felsenfeld et al., 1957). Finally, many proteins bind in the major groove, and the "structural" waters appear to be an important component of the protein-DNA interface (Otwinowski et al., 1988; Shakked et al., 1994).

A number of computer simulations of DNA hydration have been performed starting with pioneering works by Clementi and Corongiu (reviewed in Clementi (1983)), and up to recent times (McConnell et al., 1994; Cheatham and Kollman, 1996; Yang and Pettitt, 1996). These studies use both MC simulation techniques, (Clementi, 1983; Poltev et al, 1988; Subramanian and Beveridge, 1989; Eisenhaber et al., 1990; Teplukhin et al., 1991; Vovelle and Goodfellow, 1993 and references therein) and molecular dynamics, MD (Seibel et al., 1985; Swaminathan et al., 1991; Chuprina et al., 1991; Miaskiewicz et al., 1993; Fritsch et al., 1993; McConnell et al., 1994). Hydration free energies were also evaluated for DNA, based on calculation of the solvent accessible surface areas (Raghunathan et al., 1990, and references therein). In addition, the potential-of-mean-force approach was developed recently for computing the hydrophilic hydration of DNA (Hummer and Soumpasis, 1994; 1995).

The main emphasis of these studies was on the hydration shell in the minor groove, in particular on the hydration spine. E.g., our earlier results (Poltev et al., 1988; Teplukhin et al., 1991) indicate that this spine is formed to a significant extent only in A-tracts with a narrow minor groove, while in the canonical DNA with "standard" groove sizes the hydration shell is represented by "water strings" (Prive et al., 1987). Other authors refer to computational evidence for the formation of the hydration spine in the canonical B form as well (Subramanian and Beveridge, 1989; Hummer and Soumpasis, 1994), although no X-ray data on such a water pattern exist for this case.

Results of computations for the major groove are less explicit, in accord with the X-ray and NMR data indicating a more disordered state of water shell in the major groove (Prive et al., 1991; Kubinec and Wemmer, 1992; Liepinsh et al., 1992; Fawthrop et al., 1993). The role of specific hydration patterns in the major groove in stabilizing a particular conformation of DNA in solution, to the best of our knowledge has never been discussed. Among very few detailed results published so far, we mention the "regions of high water density ... between consecutive adenines and at the center of base-pair steps" (Hummer and Soumpasis, 1994). Similarly, when analyzing the B' form, we found earlier that the water molecules bridging N7 and H(N6) atoms of the two adjacent adenines often make a third hydrogen bond with an N7 atom of the adenine whose H(N6) takes part in this bridging (see Figure 1 in Poltev et al., 1988). These water molecules can also be treated in a spine-like manner, by analogy with the minor groove. This possibility was not systematically analyzed, however. The present study was undertaken to fill this gap.

Selecting the method for simulating the DNA hydration, we took into account the following considerations. In all Monte Carlo (MC) studies published so far, hydration of DNA has been simulated either for the crystal and fiber structures, or for the energy minimized DNA conformations. This approach is disadvantageous, however, since the DNA flexibility is ignored. On the other hand, the MD simulations of flexible oligonucleotide duplexes were performed by several groups of authors (Seibel et al., 1985; Zielinski et al, 1988; Miaskiewicz et al., 1993; Fritsch et al., 1993), but recently Beveridge and his colleagues concluded that "MD on DNA must be carried out for at least an order of magnitude longer than previously expected, and perhaps even longer" (McConnel et al., 1994). Although remarkable progress has been achieved in this area during the last years, and now the overall behavior of the double helix is predicted quite realistically, the details of the DNA structure are not entirely consistent with the experimental data. For example, the calculated equilibrium twisting angle for B-DNA is by several degrees less than the solution measurements (Cheatham and Kollman, 1996), so the MD-simulated B-form has certain features usually inherent in A-form. Moreover, the longest nanosecond-scale MD simulations published recently (Cheatham and Kollman, 1996; Yang and Pettitt, 1996) reveal essential force field dependence of the resulting DNA structure (B-like and A-like structures respectively).

Apparently, the unrestrained MD and/or MC simulations, which would reproduce reliably the experimentally observed difference in DNA conformation at 5° and 35°C, still remain a remote perspective. In this situation we have chosen an alternative "mixed" approach: to analyze by MC simulations the hydration of a multitude of rigid DNA conformations representing the ensemble of DNA structures in solution under different conditions. Initially, these DNA duplexes were generated in the course of MC simulations *in vacuo* (Zhurkin et al., 1990, 1991), with and without NMR restraints obtained at low temperature (Nadeau and Crothers, 1989). Thus the temperature dependent variability of the AT-containing duplexes was implicitly taken into account. In addition, we decomposed the energy of the DNA-water interaction into several terms, such as the energies of hydration of adenines and thymines in the minor and the major grooves. This was done to analyze in detail the energetic and the structural features of the water shell for different DNA conformations, with the aim of elucidating the role of hydration in stabilizing (or destabilizing) the DNA curvature at different temperatures.

Specifically, we are addressing the following questions: Is "the ensemble of states associated with the bent duplex" (cited from Park and Breslauer, 1991) more preferable than the "premelted" one in terms of hydration and, if the answer is positive, what are the structural aspects of this preference?

## METHODS

### DNA configurations

Various conformations of the DNA decamer duplexes $(A_5T_5)_2$ and $A_{10}:T_{10}$ were generated using the program *DNAminiCarlo* (Ulyanov et al., 1989; Zhurkin et al., 1991), enabling one to minimize DNA energy and perform MC simulations. For the low temperature structures, at ~5°C, the constraining pseudoenergy term was added, corresponding to the NMR restraints on the H2(Ade)···H1' distances in the minor groove (Nadeau and Crothers, 1989; Zhurkin et al., 1991). The high temperature structures (35°-40°C) were calculated without any restraints (Zhurkin et al., 1990), see Figure 1, B' and B°.

Correspondingly, two sets of $(A_5T_5)_2$ duplexes were chosen for analysis of their hydration shells (with and without NMR restraints). Each of them consists of the "averaged", energy minimized, and two "instantaneous" conformations. In all, eight $(A_5T_5)_2$ duplexes were analyzed. To define the "averaged" DNA structure, the generalized coordinates of bases and sugar rings (Dickerson et al., 1988; Zhurkin et al., 1991) were averaged over the Markov chain (for each base-pair separately). Then the geometry of the sugar-phosphate backbone was found using the chain-closure algorithm (Zhurkin et al., 1978). All the considered $A_5T_5$ structures are *irregular* in a sense that different AA:TT steps have different conformations, and thus we can elucidate the dependencies of their hydrations on the local DNA geometries. In addition to AA:TT, these decamers contain the AT:AT and TA:TA dimers, so we are able to estimate the effect of sequence on the DNA hydration.

The $A_{10}$:$T_{10}$ duplex was represented by three averaged regular conformations -- again, obtained with and without the NMR restraints -- and also by the conformation averaged over the AA:TT steps in the B-DNA crystal structures (Gorin et al., 1995), see Figure 1, $B^x$. Analysis of the three *regular* decamers $A_{10}$:$T_{10}$, each of which has ten identical AA:TT steps, allows us to get a more reliable statistics for the hydration of the AA:TT dimeric steps, and to estimate both the mean values and the fluctuations of the hydration characteristics.

Note that the B° conformation calculated without NMR restraints to mimic the high-temperature structure of *poly(dA):poly(dT),* has a positive roll angle similar to the high resolution NMR structures observed for several mixed sequences (Ulyanov and James, 1995). In this sense, B° form shown in Figure 1 can be considered as an analogue of the "averaged mixed sequence" in solution.

**Hydration shell simulations**

Standard Metropolis procedure (Metropolis et al., 1953) was used to generate Boltzmann *NVT* ensemble of water structures around a rigid DNA duplex as described earlier (Poltev et al., 1988; Teplukhin et al., 1991). To simulate hydration of the irregular DNA structures, the periodic boundary conditions along the curved DNA helical axis were applied (Zhurkin et al., 1991). The unit cell contained DNA decamer duplex and 250 or 500 water molecules. This amount of waters is sufficient for obtaining reliable data on base hydration, as the principal results related to the bases were found to be essentially the same for both 250 and 500 waters (Table 1). To maintain the constant volumes of the systems, the distances from the waters to the geometrical centers of the closest DNA structural units (base, sugar or phosphate) were limited by 9.5 Å for the system with 250 waters, and by 11 Å for 500 waters. To achieve equilibrium, the Markov chains consisting of 50,000 trials per water molecule have been generated. Then the hydration characteristics were averaged over 500,000 trials per water, which corresponds to $1.25$-$2.5 \cdot 10^8$ system configurations.

The energies of water-water and water-DNA interactions were calculated using the atom-atom potentials specially adjusted for nucleic acids hydration (Poltev et al., 1984). These potentials realistically reproduce the bulk water properties and the hydration energies for the nucleic acid bases (Poltev et al., 1996). Since the present study is aimed mainly at elucidating the details of the sequence-specific water-base interactions, the cations were excluded from the consideration, and the phosphate groups were neutralized (Kollman et al., 1982; Zhurkin et al., 1982).

When estimating the contributions of the minor and the major grooves in the hydration energy, only the waters located no further than 7.5Å from the center of the nearest base were taken into account. Whether a water molecule belongs to the minor or major groove was determined by which base atom is the nearest to it. The individual energy terms corresponding to interactions of each base, sugar and phosphate with waters were calculated as well (to make the structural units approximately neutral, O3' atoms were assigned to sugars, and O5' atoms to phosphates).

**Water structure analysis**

To analyze the water configurations accepted in the course of MC run, the DNA-water and the water-water hydrogen bonds were identified using the following geometric criteria (Poltev et al., 1984): the distances O···H (N···H) and O···O (N···O) should be no more than 2.4 and 3.2Å, respectively. Each DNA hydrophilic atom was characterized by a "hydration index" calculated as a number of the water molecules hydrogen bonded to this atom.

To describe a water configuration, the following hydration patterns were used:
*(i) water bridges* (water molecule making two hydrogen bonds with DNA atoms simultaneously; bifurcated hydrogen bonds were treated separately), *(ii) water "tridents"* (three hydrogen bonds with the DNA atoms per one water molecule), *(iii) two-water bridges* (two molecules forming hydrogen bonds with DNA are H-bonded), and *(iv) three-water bridges* (when two of them forming hydrogen bonds with DNA are H-bonded with the same third molecule). The probability of occurrence of each hydration pattern was calculated as the ratio of the number of configurations in which the pattern was revealed to the total amount of generated configurations.

# RESULTS AND DISCUSSION
**General features and energetics**

The conformations chosen to represent the B'-form generated with the NMR restrictions at low temperature, proved to be more preferable than the "premelted" B° conformations from the viewpoint of the hydration energy. This difference is readily illustrated by the data for the two $A_{10}:T_{10}$ duplexes (Table 1), where it is evident that the preference of the B' form hydration originates mainly due to the favorable hydration energy of bases, $E_{base}$.

To make sure that the results do not depend on the size of the simulated water shell, we analyzed separately hydration of the duplexes with 250 and 500 water molecules. The dependence of the hydration energy on the size of the water shell was found to be different for the bases, sugars and phosphates. As expected, the strongest effect was observed for the phosphate groups positioned on the periphery of the double helix: their hydration energy decreased by as much as 8-10 kcal/mol when the amount of waters was doubled ($E_{phos}$ in Table 1). In contrast, hydration of the bases and the sugar rings located closer to the DNA axis, decreased only by 2-3 kcal/mol ($E_{base}$ and $E_{sug}$ in Table 1). Nevertheless, the total difference in the hydration energy between the B' and B° forms proved to be practically the same for the both water shells: $\Delta E_{total} \sim 4$ kcal/mol, and in both cases this difference is related mainly to the hydration of bases (Table 1). Thus, our main result on the preferential hydration of the B' form does not depend on the particular choice of the hydration shell.

Notice that the water-water interactions, as well as the intra-molecular DNA-DNA interactions, are less favorable for the B' conformations (Table 1). However, a considerable gain in the energy of the water-DNA interactions is twice as much as the loss in the energy of water-water and DNA-DNA interactions, so the total energy of the system (DNA+water) is lower for the B' conformation compared to B° (see Table 1). Importantly, results of our direct MC simulations are in general agreement with the data reported by Raghunathan et al. (1990), who calculated the solvent accessibility surface areas for various DNA conformations and concluded that the solvation energy of *poly(dA):poly(dT)* is lower when it has a wider major groove.

Consider hydration of adenines and thymines separately. As follows from Table 2, the water-adenine interactions provide basic impact into this gain: compare E(A) and E(T) values. The more detailed analysis revealed that hydration of the adenine N7 differs drastically for the B' and B° conformations, while the differences in hydration of the N3 and O2 atoms are less pronounced (the "hydration index" of N7 varies by ~0.4 compared to 0-0.1 for N3 and O2, see Table 2). Next, we treated the base hydration in the minor groove separately from that in the major groove. We found that it is the energy of the adenine-water interactions in the major groove, $E_M(A)$, that is the bulk of the DNA hydration energy increase on DNA "premelting", or B' → B° transition (Table 2, lines 3-6). The change in the hydration of the minor groove is substantially less pronounced: compare the energy difference for the minor groove, $\Delta E_m(A)+\Delta E_m(T) \sim 1$ kcal/mol, with the corresponding value $\Delta E_M(A)+\Delta E_M(T) \sim 4$ kcal/mol for the major groove (Table 2).

The above results for the regular $A_{10}$:$T_{10}$ decamers are entirely consistent with the data obtained for the irregular $A_5T_5$ conformations. This point is illustrated numerically for the two "average" structures, B' and B° (Table 3): the hydration is again more advantageous for the B' conformation, and this advantage is achieved mainly due to the hydration of bases, in particular of adenines in the major groove (compare the values $E_{base}$, $E(A)$ and $E_M(A)$ in Table 3). In addition, the same trend was found when various conformations of $A_5T_5$ were considered, and the hydration energies were analyzed as functions of the groove sizes (Figure 2).

We use the distance between the closest sugar ring oxygens O4' of the opposite strands as a measure of the minor groove width (Figure 3a; see also Prive et al., 1991). Such a choice provides the strongest correlations with various hydration characteristics in comparison with the interphosphate distance (see Figure 4 and below).

The irregular $A_5T_5$ duplexes considered here represent a very wide range of the minor and major groove widths, both of which vary nearly twofold (Figure 2). Nevertheless, the general tendency is clear-cut for the whole multitude of DNA structures: the hydration energy of bases has only slight dependence on the minor groove width (Figure 2a), whereas the energy decreases rather steeply with the major groove widening, especially in the case of adenine (Figure 2b). In addition, there is one more feature distinguishing the minor and major grooves. In the case of the major groove, hydration energies of both the adenines and the thymines decrease monotonically (Figure 2b), so when the total hydration energy in the major groove is considered, $E_M(A)+E_M(T)$, it gives a decrease of ~1 kcal/mol per 1Å of the groove width (data not shown). In contrast, as far as the minor groove is concerned, adenines and thymines reveal opposite trends: in the case of thymines, the hydration energy gradually increases with the widening of the groove, but for adenines the energy decreases overall (Figure 2a). As a result, the sum of the hydration energies for adenine and thymine in the minor groove remains nearly constant (data not shown).

Notice that the hydration energy of adenines in the minor groove varies non-monotonically: hydration is the weakest for the groove width ~7.5Å which roughly corresponds to the "averaged" X-ray structure of the A-tract (Figure 2a, lower panel). This effect is relatively weak, however, and further analyses will be necessary to clarify this issue.

Finally, based on all the data presented above, both for regular $A_{10}$:$T_{10}$ and irregular $A_5T_5$ decamers, we conclude that the B' conformation of the A-tract is hydrated more strongly than the B° form. Thus, the first question raised in the Introduction is resolved positively, namely, the low temperature conformation B' does appear to be more suitable for hydration compared to the high temperature "premelted" conformation B°. The next question related to the correlation between the energetics and the structural aspects of DNA hydration, is addressed below, where the minor and the major grooves are analyzed separately in more detail.

**Minor groove hydration**

The minor groove of the AT-containing DNA fragments contains four hydrophilic centers per nucleotide pair, all of them hydrogen bond acceptors. Two of them belong to a base pair, adenine N3 and thymine O2, and the other two are the sugar oxygens, O4' (Figure 3a). As is well known, the N3/O2 atoms are located nearly symmetrically with respect to the dyad axis of the AT-pair (Seeman et al., 1976). This allows one to consider all the AT-containing dimeric steps within the same framework.

As far as the inter-atom distances are concerned, numerous one- and two-water bridges between the hydrophilic centers from the same pair, or from the two adjacent pairs may be formed. Most often, however, only three main patterns of the water structure were observed, in each of which the waters are directed "across" the groove, as shown in Figure 3a. The patterns are denoted *m1*, *m2* and *m3*, where *m* stands for the minor groove (Figures 3b-3d and Table 4):

<u>*m1*</u>: the inter-strand one-water bridge between the two N3/O2 atoms of the neighboring base-pairs -- the first layer of the so called "hydration spine" (Drew and Dickerson, 1981). These atoms are N3 and O2 in the case of AA:TT step, two atoms O2 in the AT step, and two atoms N3 in the TA step (Figure 3a).

<u>*m2*</u>: the inter-strand two-water bridge between the same atoms N3/O2 as in above case (Figure 3b); see also Figure 7b in Goodsell et al. (1994).

<u>*m3*</u>: the intra-strand water bridges between the N3/O2 atom of a base and the sugar ring oxygen O4' from the 3'-adjacent nucleotide, denoted as "strings" by Prive et al. (1987).

As follows from our calculations, the probabilities of formation of the patterns *m1* to *m3* depend directly on the minor groove width. For the narrow minor groove the pattern *m1* is predominant, but when the groove is widened, the *m2* and *m3* patterns occur more frequently. First, we consider hydration of the AA(AA:TT) dimeric steps, and then discuss the AT and TA steps.

<u>*Pattern m1, hydration spine*</u>
Probability of occurrence of the *m1* pattern is shown in Figure 4a for various conformations of $A_5T_5$. This probability clearly anticorrelates with the groove width. The three columns of data correspond to the three inter-atomic distances that could be used as a measure of the minor groove size. The correlation is the strongest for the sugar-sugar (O4'⋯O4') inter-strand distance: the correlation coefficient is -0.91, while for the base-base (N3/O2⋯N3/O2) and the phosphate-phosphate distances the coefficients are -0.80 and -0.83 respectively. Therefore below we describe the minor groove width in terms of the sugar-sugar distance.

There is a noticeable difference in hydration of adenines and thymines: when the "spine" water molecule is built into the narrow minor groove (not wider than 7 Å), its hydrogen can be bifurcated between O2(Thy) and the neighboring O4'(sugar). Such a hydrogen bond occurs with probability as high as 60%, while the hydrogen bonds bifurcated between N3(Ade) and the corresponding O4' remain very rare (~5%). The frequent bifurcation in the case of thymine is likely to be related to the shorter distance between the two donors: O4'⋯O2(Thy) is 3.3 Å compared to 3.6 Å for O4'⋯N3(Ade).

The double-layered hydration spine observed initially in the X-ray structure (Drew and Dickerson, 1981) was detected in our simulations as well; it occurs mostly in the narrowest parts of the minor groove. Probability of the two spine waters to be bridged by a third water molecule (so that the structure with six simultaneous hydrogen bonds is formed) rises from 10% to more than 30% as the groove narrows from 8 Å to 6.5 Å (data not shown). The data for the $A_{10}{:}T_{10}$ decamer agree quantitatively with the above results for $A_5T_5$. When the minor groove is compressed from 8.7Å to 6.4Å, the probability of *m1* increases from 21% to 92%, and that of the spine increases from 0% to 22% (see the first two lines in Table 4). Notice that the "average" X-ray structure of $A_{10}{:}T_{10}$ is similar to the "NMR restrained" structure with very narrow minor groove in a sense that the *m1* probability is very high in both cases, and the hydration spine is nearly as stable (compare B' and $B^x$ in Table 4).

*Pattern m2, two-water bridges*
When the minor groove width is larger than 8.5 Å, the two-water bridges dominate. The probability of their formation raises from 20% for the groove width of 8 Å to 50-60% for the 9-10 Å groove (see the data for $A_5T_5$ in Figure 4b). Again, the correlation coefficient is the largest, 0.81, when the "sugar-sugar" distance is used as a measure of the groove width. Similar data were obtained for the $A_{10}{:}T_{10}$ decamer: the probability of *m2* increases from 0% to 3% to 31% in the order B' - $B^x$ - B° (Table 4).

*Pattern m3, intra-strand sugar-base water bridges (strings)*

These "strings" demonstrate nearly the same tendency as the two-water bridges (type *m2*). In the case of $A_5T_5$ decamer, the probability of formation of the "strings" is ~15% for the 8 Å groove width, and increases up to ~40% when the groove widens to 10 Å (data not shown). For $A_{10}$:$T_{10}$ this probability increases from 0% for the B' form to 10-20% for the B° form (Table 4).

Notice that there are practically no water bridges directed "*along*" the minor groove: the N3..N3 and O2···O2 bridges are formed with negligible probability (Table 4). The N3···N3 and O2···O2 distances are quite suitable for the formation of hydrogen bonds with water, however (4.0-4.3Å). Therefore, there should be some steric reason to exclude stabilization of such water bridges. Indeed, we found that the N3···N3 and O2···O2 water bridges are stereochemically plausible and energetically favorable *per se* (though less favorable than the N3···O2 bridges by 2-5 kcal/mol depending on the DNA conformation). But when two strings of such N3···N3 and O2···O2 bridges were built into the minor groove, this caused repulsion between the bridges and between the bridges and the sugar O4' oxygens (data not shown). Thus, we suggest that an extremely low occurrence of the water bridges directed *along* the minor groove is a consequence of the interactions between the waters, N3/O2 atoms on the "floor" of the groove, and the O4' oxygens on the sugar-phosphate "walls".

Above, we have analyzed the minor groove hydration patterns for the AA:TT step in both $A_5T_5$ and $A_{10}$:$T_{10}$ decamers. Now we proceed to the comparison between the AA, AT and TA steps; only the $A_5T_5$ decamer will be considered.

**Weak sequence-dependence of the minor groove hydration**

It follows from our data that a single set of structural patterns (*m1*, *m2* and *m3*) can be used to describe hydration of all the AT-containing steps. In other words, the influence of the nucleotide sequence on the water structure in the minor groove of AT-containing DNA is indirect -- apparently, this occurs via the groove size.

In Figure 4a the probability of formation of the *m1* bridge is shown as a function of the groove size. This is nearly the same for all three dimers: AA, AT and TA, which illustrates the point -- it is not the sequence *per se* that matters, but rather the distance between the donor groups in the minor grove. This is further evident from Figure 5 where the same probability is shown *versus* the sequence. On the one hand, we see here what was mentioned before, namely, that the occurrence of *m1* is inversely proportional to the groove size. Notice, for example, a strong anti-correlation between the probability of *m1* and the sugar-sugar distance for the irregular "instantaneous" B' configuration (broken lines in Figures 5a and 5b). On the other hand, the groove size turns out to be the smallest for the AATT quartet, and the largest for the TTAA quartet, for both B' and B° structures (Figures 5a and 5c). As a consequence, the probability of *m1* is usually the largest for AATT, and the smallest for TTAA (Figures 5b and 5d).

The typical "frozen" water structures for the AATT and TTAA tetramers are presented in Figure 6a. They were obtained after MC runs at 5K starting from the final points in the Markov chain generated at 300K. These "F-structures" (Malenkov, 1985) correspond to the local minima of the total energy of the system. Notice that in agreement with the X-ray data the narrow groove of AATT is hydrated by a single hydration spine, whereas the enlarged groove of TTAA is hydrated by a double ribbon of the sugar-base strings (Prive et al., 1987) and two-water bridges (Goodsell et al., 1994).

Nevertheless, the above said does not imply that the probability of occurrence of the one-water bridge *m1* in TTAA tetramer is necessarily small. Similarly, this occurrence is not necessarily close to 100% in AATT and AAAA. This probability depends on the particular DNA conformation: for example, in the TTAA fragment it varies from 20% to 70% for the structures obtained with the NMR restraints and therefore having relatively narrow minor groove (Figures 5a-b). Our results are consistent with the recent findings by Goodsell et al. (1994). These authors detected the *m1* pattern in the TA step with the relatively narrow groove, and the two-water bridge *m2* in the TA step with the wide groove (see Figure 7 in Goodsell et al., 1994). This is exactly what we found in MC simulations.

Thus we conclude that the particular structure of the water shell in the minor groove of the AT-containing DNA is primarily sequence-independent. The narrow minor groove (O4'···O4' distance less than 7.5 Å) is preferentially hydrated by a single hydration spine, whereas the wide minor groove (9 Å and more) is filled mainly by the two-water bridges or strings (*m2* and *m3* in our notations). Both conclusions are true for any of the dimeric steps: AA, AT and TA.

Based on this uniformity, we can suggest a putative explanation to the unexpected observation that the structure of the minor groove hydration shell is relatively insignificant for stabilizing one B-DNA structure compared to another (see the previous section and Figure 2a). Comparing the *m1*, *m2* and *m3* patterns (Figures 3 and 6) we see that in each of them the strongly electronegative atoms, N3(Ade) and O2(Thy), are hydrated only once, independent of the particular orientations and positions of waters in the first hydration shell (see hydration index in Table 2). As a result, the energy of DNA hydration is only slightly affected by formation of the one-water spine or the two-water bridges in the minor groove. This is in striking contrast to the major groove hydration, as shown in the next section.

**Major groove hydration patterns**

There are two features distinguishing the major and the minor grooves of AT-containing DNA, which appear to be the most important for the present analysis. First, in the major groove there are both donors and acceptors of protons, localized asymmetrically with respect to the dyad axis of the AT-pair (Figure 7a). Consequently, the pattern of the hydrophilic centers is essentially sequence-dependent and for the mixed sequence, quite irregular (Seeman et al., 1976). Second, the hydrophobic major groove "walls" are located farther away from the hydrophilic base atoms compared to the minor groove, and thus these "walls" do not form a narrow cleft restraining the positions of the hydration waters. As a result, interactions of water molecules with the base hydrophilic centers appear to be much more important than interactions with the backbone.

The distances between the hydrophilic atoms on the edges of the bases are suitable for the formation of numerous intra- and inter-strand water bridges, containing one, two or three water molecules. So, it is not surprising that various relatively unstable hydration patterns were observed in the major grooves of the $A_5T_5$ and $A_{10}:T_{10}$ duplexes. Since arrangement of the hydrophilic centers in the major groove depends on the nucleotide sequence, one has to use three different sets of hydration patterns to describe the dependency of the water structure on DNA conformation in the AA, AT and TA steps.

Hydration patterns of the AA step. The most important of these are described below and displayed in Figure 7. They are denoted *M1, M2, ... M5,* where *M* stands for the major groove. Their probabilities for $A_{10}:T_{10}$ duplexes are presented in Table 5. Notice that these water bridges are directed both *along* and *across* the major groove (Figure 7a), thus making the hydration structure less ordered in general than in the minor groove (Figure 3a). Correlations of the water bridge probabilities with the major groove width are also less clear-cut compared to the minor groove. Nevertheless, several general trends can be observed:

*M1* the intra-strand one-water bridge between N7 and H(N6) atoms of adjacent adenines (Figures 7a, 7b). The *M1* pattern arises with an appreciable probability (up to 60%) in all AA steps of $(A_5T_5)_2$ decamers and in the hydration shells of all three $A_{10}:T_{10}$ duplexes (Table 5).

*M2*   the intra-adenine bridge between the two "Hoogsteen sites", N7 and H(N6) (Hoogsteen, 1959), see Figure 7a.  The bridge *M2* is a feature intrinsic to an individual adenine; it was predicted in the course of calculations of hydration shells of the isolated adenines and AT-pairs (Poltev et al., 1992). Nevertheless, its probability depends on the DNA conformation, generally being higher for wider major grooves:  the *M2* occurrence is 21% , 31% and 38% in B°, B$^x$,  B' , respectively (Table 5). This water bridge was observed in at least two B-DNA dodecamers : in CGCGAATTCGCG (Drew and Dickerson, 1981; Table 2; waters #49, 66) and in CGCAAATTTGCG (Edwards et al, 1992; Figure 6; waters #27, 44); see also Figure 2 in Schneider et al. (1992) and Figure 2d in Schneider et al. (1993).

*M3*   the intra-strand bridge between two N7 atoms of adjacent adenines (Figure 7a).  If these three one-water bridges (*M1*, *M2* and *M3*) are formed by the same water molecule, it is denoted as a "water trident", i.e. a water molecule hydrogen bonded to three hydrophilic atoms of DNA simultaneously (N7 and H(N6) of the 3'-end adenine, and N7 of the 5'-end adenine), see Figure 7d. The probabilities of *M3* bridges and those of the tridents differ substantially for the B° and B' conformations of $A_{10}$:$T_{10}$, so the high probabilities of these hydration patterns can be considered as distinctive features of the conformations with a widened major groove and a compressed minor groove, see Table 5.

Note that a water trident bound to the AG:CT dimeric step was observed crystallographically in the *trp* repressor/operator complex (Shakked et al., 1994), see Figure 7e.  The geometries of the calculated and the observed tridents are remarkably similar in general, although orientations of the water molecules are different in the two cases: in the AG-bound water one proton and oxygen are bound to N7 and H(N6) of the 5'-end adenine, and the second proton to O6 of the 3'-end guanine (Figure 7e), whereas in the AA step only one proton is bound to the 5'-end adenine (Figure 7d). This difference is naturally explained by the presence of an acceptor of protons in the 6-th position of guanine, and a donor of protons in the 6-th position of adenine.

*M4*   inter-strand water bridge across the major groove connecting O4(Thy) and H(N6)(Ade) from adjacent AT-pairs (Figure 7a).  The *M4* bridge occurs most frequently (up to 60%) in the narrowest portion of the major groove (~13 Å), and does not appear if the groove width is more than 16.5 Å. This hydration pattern was reported for the CGCAAATTTGCG dodecamer (Edwards et al, 1992; Figure 6; water #38).

Notice the opposite tendencies in behavior of the *M4* bridge, on the one hand (the probability is higher when the groove is narrow), and the *M3* bridge and the trident, on the other hand (the probability is higher when the groove is wide); see Table 5. The structural reason for this difference is apparently related to the orientation of the H(N6) proton (see Figures 7a - 7d). Proton

H(N6) is turned away from O4, so an increase in the groove width, and simultaneously in the O4⋯N6 distance, separates the atoms H(N6) and O4 too much, and thus makes the *M4* bridge unstable (Figure 7c). On the contrary, a similar increase in the N6⋯N7 distance makes formation of the trident more favorable, because H(N6) proton is directed toward the two acceptors N7 (Figure 7d). This simple stereochemical factor, the orientation of the adenine amino-group, appears to be directly related to the preferential formation of water tridents for the A-tract conformations with wide major grooves.

*M5* intra-strand bridge between two O4 atoms of adjacent thymines (Figure 7a). The probability of the *M5* pattern is relatively low for all conformations and does not exceed 35%. For the regular duplexes presented in Table 5, this probability is less than 15%.

Among the two-water bridges arising with the probability up to 40% it is worth mentioning the O4⋯N7 bridge between adjacent AT base-pairs, which was also observed in the X-ray CGCAAATTTGCG dodecamer (Edwards et al, 1992; O4(Thy8) and N7(Ade16) in Figure 6).

Hydration patterns of the TA step.
The patterns formed most frequently are the single-water and the two-water bridges across the groove between the O4(Thy) atoms from the opposite chains (wavy line in Figure 7a, *below*). The one-water bridge probability grows from 70% to 90% as the O4⋯O4 distance increases from 3.2 Å to 4.2 Å (meanwhile the major groove width increases from 14.3 to 15.5 Å). Then this probability drops down to 10% as the distance O4⋯O4 becomes greater than 5 Å (the major groove becomes wider than 17.5 Å). Meanwhile, the two-water bridge probability grows from ~10% up to ~70%.

Importantly, the probabilities of these bridges depend on the O4⋯O4 distance more strongly than on the groove width. This is consistent with the notion that the major groove is relatively wide and the hydration pattern is governed by the local atom-atom distances rather than by the groove size. Note that the one-water bridge O4⋯O4 was observed in the crystal structures of CCATTAATGG and CGATTAATCG duplexes (Goodsell et al., 1994; Figure 7), whereas the two-water bridge between the same atoms was detected both in free DNA and in the *trp* repressor/operator complex (Shakked et al., 1994; Figure 3b, waters W5 and W5').

Hydration patterns of the AT step.
There are no highly probable bridges for this step. The one-water and two-water intra-strand bridges between the adenine N7 and the adjacent thymine O4, as well as the inter-strand bridges between H(N6) of two adenines are formed with the probability of 10-30% (wavy lines in Figure 7a). Notice that the N7···O4 two-water bridge detected in our calculations was revealed recently in the X-ray structure of the *Hin* recombinase-DNA complex (Feng et al., 1994). Remarkably, both these waters participate in the protein-DNA recognition and may be stabilized by the protein.

In conclusion, the water shell in the major groove of AT-containing DNA reveals much greater variability, and the stability of each particular hydration pattern in the major groove is noticeably less than in the minor groove. This difference is related to the major groove features mentioned above: the groove "walls" are mainly hydrophobic, and these "walls" are located farther away from the base hydrophilic centers compared to the minor groove. Thus the "walls" do not create a limited space for the hydration waters; as a consequence, the local changes in interactions between waters and it's nearest hydrophilic atoms play the most critical role.

GC-containing steps. Detailed study of the GC-containing dimers lies beyond the scope of this paper, and will be presented elsewhere. Here we wish to mention briefly that there is a principal difference between the AT and GC pairs in terms of suitability for the water bridges and cations. Notice that the protons H(N6) of adenines which are not involved in the Watson-Crick interactions (shown as black triangles) are *directed toward* the closest acceptors of protons, N7(Ade); such orientations of these amino-protons facilitate formation of the water tridents (Figures 7d, 7e). Whereas in GC pairs, the analogous non-WC protons H(N4) of cytosines are *turned away* from the acceptors O6(Gua), which is likely to diminish the stability of putative water tridents in GG, GC and CG steps (Figure 7f). In particular, the trident observed in GC dimeric step of the B-DNA dodecamer and shown schematically in Figure 7f, is characterized by increased distances between the water oxygen and atoms O6/N4/N7 (these distances are 3.5-3.8 Å, see water #10 in Figure 10 and Table 2 in Drew and Dickerson, 1981).

One more difference between the AT- and GC-containing sequences in the major groove, is a compact location of the electronegative N7/O6 atoms in GC-fragments, a feature following immediately from Figure 7f, especially for the GC and GG steps. The negatively charged atoms N7(Gua) and O6(Gua), located next to each other, are likely to serve as binding sites for the divalent cations, such as $Mg^{++}$ and $Zn^{++}$, which, in turn, may cause base-pair rolling toward the major groove and enhance curvature in the GC-rich DNAs (Shlyakhtenko et al., 1990; Nickol and Rau, 1992; Brukner et al., 1994; see also Figure 5b in Olson and Zhurkin, 1996).

**Hydration energy and water tridents at the adenine N7**

In spite of the relatively weak correlations between the water bridge probabilities and the major groove size, several remarkable relationships were observed in the previous section. Namely, increase in the major groove size is linked with the increase in frequencies of the water tridents and the intra-strand M2 and M3 bridges from 10-20% to 25-40% (see the data for two $A_{10}$:$T_{10}$ structures in Table 5). These hydration patterns involve interaction with the N7 atoms of adenines (Figure 7d). Therefore, we analyzed in more detail how the N7 hydration depends on the groove size for various $A_5T_5$ decamers.

For this purpose the so-called "hydration index" of N7, or the average number of the hydrogen bonds between N7(Ade) and waters, was considered (Figure 8). The hydration index increases by ~0.5 when the double helix undergoes a transition from the B/A-like structure with a positive roll and compressed major groove (Figure 1, B°) to the C-like structure with a negative roll and wide major groove (Figure 1, B'), see Figure 8a. The correlation between the N7 index and the groove width is relatively weak, and the points are noticeably scattered in Figure 8a. Nevertheless, this modest increase in the average number of waters bound to N7(Ade) is accompanied by a significant decrease in the energy of hydration by ~6 kcal/mol (Figure 2b).

The direct connection between the hydration energy of adenines in the major groove and the N7 index is obvious in Figure 8b: the correlation coefficient in this case is -0.94 versus 0.61 for Figure 8a. For the other two hydrophilic centers, H(N6)(Ade) and O4(Thy), this correlation is weaker (data not shown). Thus, it is conceivable that the energy of hydration is related to the interactions of waters with N7(Ade).

This result was obtained for various configurations of the $A_5T_5$ decamer. Similar conclusion follows from the data for $A_{10}$:$T_{10}$ (Table 2): the B° → B' transition corresponds to an increase in the hydration index of N7 by 0.36, whereas the same indexes for H(N6) and O4, taken together, increase by 0.25 only. The B° → B' transition decreases the adenine hydration energy by 3.4 kcal/mol (Table 2). When this energy difference was decomposed into the direct hydrogen-bonded interactions and the non-specific water-base interactions without hydrogen bond formation, the latter proved to contribute ~1 kcal/mol, whereas the water tridents comprised the main contribution, about 2.5 kcal/mol.

The snapshots of the hydration shells in the major groove of A-tract are given in Figure 6b. As in the case of the minor groove (Figure 6a), these "frozen" configurations obtained at 5K correspond to the local minima of hydration energy. Three rows of waters can fit into the increased major groove of the B' form, whereas in the B° form with the "normal" major groove there is enough space only for two rows of water molecules.  As a result, the hydrophilic atoms N7 are more easily accessible for waters in the B' form compared to the B° form.  These snapshots illustrate the same dependence of the N7 index on the major groove size as shown in Figure 8a.

The difference in hydration of adenines in the two DNA forms is shown in more detail in the skeleton representation (Figure 9).  In the B' form the adjacent water tridents interact in a regular fashion with the three consecutive adenines.  By contrast, the hydration shell of the B° form is disordered, and the water chains are much less regular; they are represented mainly by two-water bridges and the one-water bridge *M1*.  The principal point here is that in the case of the B' form the most electronegative N7 atoms are hydrated twice, as opposed to the B° form where each N7 atom is hydrated only once.  This appears to be the main origin of the energetical advantage of the trident scheme.

The main result of this section can be summarized in the following way.  The hydration energy of DNA in the major groove strongly correlates with the N7(Ade) hydration index: the larger the number of water molecules hydrogen bonded to N7 atoms, the lower the energy (Figure 8).  The number of waters bound to N7 atoms is increased when the adenines are separated in the major groove, and the groove itself becomes wider (Figure 6b).  This occurs, for example, when base pairs are rolled toward the minor groove and inclined as in the B' form (Figure 9).  In turn, the roll angle is directly linked with the DNA bending and curvature.  Thus, the B' form of the A-tracts, attributed to the ability of the A-tracts to cause the DNA curvature, is characterized by a propensity for stronger hydration of adenines in the major groove.

## CONCLUSION

Results of MC simulations of AT-containing DNA duplexes suggest that the adenine-water interactions in the major groove help stabilize the A-tract conformations with the narrowed minor groove and widened major groove.  These conformations are the result of a delicate balance between the stacking and hydration interactions, which can be disrupted by rising temperature, binding of netropsin-like ligands in the minor groove, exclusion of water from the major groove, or by other perturbations changing the water network.  Thus, in the structural interpretation of experimental data on DNA conformational transitions in general, and the DNA bending/curvature in particular, the major groove hydration should be taken into account in addition to the well known Drew-Dickerson hydration spine in the minor groove.

Considering structural waters in the major groove helps explaining at least three sets of data, which otherwise remain at odds with the current paradigm, suggesting that minor groove hydration is the main factor responsible for stabilization of the B' form (or B'-like form) of A-tracts in solution (Chuprina, 1985, 1987). First, changes in the hydration of the major groove provide an additional source of immobilized water molecules accounting for a substantial molar volume increase detected upon binding of netropsin in the minor groove (Marky and Kupke, 1989). Two more examples are related to the decrease in DNA curvature caused by substitution of adenine by inosine (Diekmann et al., 1987) and by 7-deazaadenine (Seela et al., 1989). In both cases the acceptor/donor pattern of N7 and H(N6) groups is distorted, which leads to disruption of the water tridents (Figure 7).

We interpret the results of MC simulations (Figure 8) to mean that the water tridents bound to the N7 and H(N6) atoms of the adjacent adenines, act as "strings" (or "wedges") pushing two adenines apart from each other and creating the negative roll in A-tracts (Figure 9, B'). Such water tridents are impossible when the H(N6) groups are substituted by O6 in inosines, or the N7(Ade) atoms are substituted by H(C7) groups in 7-deazaadenines. As a consequence, the roll angle in the "substituted A-tracts" becomes less negative, conformation of these tracts becomes closer to that of a "mixed" sequence, and the DNA curvature is decreased.

Thus, we suggest that the structural waters play an important role in determining the sequence-dependent conformational preferences of the DNA duplex. In particular, various water bridges have to be considered in addition to the interactions between the exocyclic groups (Diekmann et al., 1987; Hunter, 1993; Sponer and Kypr, 1994; Gorin et al., 1995). Based on the results presented here (see also, Teplukhin et al., 1996), we propose that major groove hydration is one of the main driving forces causing conformational transition in $A_n:T_n$ tracts and strong DNA curvature in solution at low temperature (Teplukhin et al., 1995). This concept, together with the "anisotropic flexible wedge model" (Zhurkin et al., 1991; Olson et al., 1993), accounts for most of the available experimental data on the temperature and sequence-dependent curvature of DNA with $A_n:T_n$ tracts. The logical scheme of the proposed concept is outlined below.

(i) DNA bending is anisotropic: it is energetically more advantageous to bend duplex toward the grooves, in the Roll direction, than in the Tilt direction (reviewed by Olson and Zhurkin, 1996).

(ii) The double helix is asymmetric with respect to the Roll angle distortions, and this asymmetry is sequence-dependent (Ulyanov and Zhurkin, 1984; Srinivasan et al., 1987; Sarai et al., 1989; Olson et al., 1993). In particular, the AA:TT step reveals tendency toward the negative rolls, whereas the GC-rich sequences are biased toward the positive rolls. Thus *on average,* when the DNA fluctuations are taken into account, the base-pairs in $A_n:T_n$ tracts are expected to be effectively rolled into the minor groove compared to "mixed" sequences. This is in qualitative accord with the well known "static AA wedge model" by Ulanovsky and Trifonov (1987), postulating the minor groove rolling in the AA:TT steps.

(iii) The estimates of this difference in roll between the $A_n:T_n$ tracts and the "mixed" sequences vary from 3° to 9° (Ulanovsky and Trifonov, 1987; Maroun and Olson, 1988; Calladine et al., 1988; Bolshoy et al., 1991; Goodsell et al., 1994; Haran et al., 1994). At the same time, the thermal fluctuation of the roll angle in DNA is about 7°, which corresponds to the persistence length of 500Å (Hagerman, 1988b; Olson et al., 1993). We see that the two mentioned values are comparable. Thus the DNA curvature in solution is a very delicate effect comparable in scale to the energy of thermal fluctuations, and as such it should depend on environment. Due to interactions with water and ions the average DNA conformation in each particular step can be easily shifted toward the positive or the negative rolls.

(iv) At low temperature (~5°C), when the enthalpy effect is dominating, we assume the $A_n:T_n$ tracts are hydrated in the energetically optimal way, with the formation of water tridents shown in Figure 9 (B'-form). As a consequence, the major groove is widened in A-tracts, and the minor groove is compressed, facilitating formation of the hydration spine (Drew and Dickerson, 1981; Chuprina, 1985). The B'-like form is stabilized with base-pairs rolled toward the minor groove similar to *poly(dA):poly(dT)* in fibers (Alexeev et al., 1987). Importantly, the mixed sequences do not have the specific N7/H(N6) pattern in the major groove (Figure 7a), thus their base-pairs are unlikely to be rolled into the minor groove under the influence of hydration. This difference in roll between the mixed sequences and the A-tracts could account for DNA curvature (Ulanovsky and Trifonov, 1987; Crothers et al., 1990).

(v) At relatively high temperature (35°-40°C), the entropy plays a more active role, and the hydration shells in both grooves are likely to be more disordered. Thus, according to our interpretation, the "canonical" B-DNA is formed at the high temperature as a result of this increased mobility of waters and disruption of the water tridents in the major groove, because the B-like form is preferable for the base stacking *per se* (like $B^x$ in crystals in which the base-pairs are nearly perpendicular to the helical axis, or like our calculated B° *in vacuo* where the base-pairs are slightly rolled toward the major groove, see Figure 1).

(vi) We suggest that the similar role in disrupting hydration patterns of A-tracts is played by various alcohols, such as EtOH and MPD (Marini et al., 1984; Sprous et al., 1995). We expect that in the presence of these nonelectrolytes the water shells in the DNA grooves are distorted and the probability of formation of the water tridents in the major groove is diminished. As a consequence, the roll of bases in A-tracts toward the minor groove would be less pronounced, the minor groove would widen, the ensemble of the A-tract structures would become closer to that for the mixed sequences, and consequently, the DNA curvature would decrease.

The last item (vi) is apparently in accord with the recent data by Dlakic et al. (1996) indicating that the A-tracts have an increased width of the minor groove in the presence of MPD (as revealed by an increase in the DNase I cleavage), whereas the GC-rich "mixed" sequences retain the same width of the groove (as follows from the same rate of cleavage). In a similar way one can account for the role of glycerol causing the DNA straightening, but not for the role of sucrose apparently increasing the DNA curvature in polyacrylamide gel (Pennings et al., 1992).

Available data do not allow us to make any conclusions on the detailed molecular mechanism(s) of the nonelectrolyte-water-DNA interactions. There are two alternative ways how a cosolvent could in principle influence the hydration shell structure in the DNA grooves:

(I) Cosolvent can rearrange the DNA-water interactions and the DNA conformation through *non-specific effects* which are being described in various terms: activity (chemical potential) of water (Malenkov et al., 1975, 1979); osmotic stress (Parsegyan et al., 1995); crowding effect (Zimmerman and Minton, 1993); preferential hydration (Timasheff, 1993). The changes in the cation screening of the phosphate groups upon addition of nonelectrolyte can also be considered as a non-specific effect (Dickerson et al., 1996).

(II) Cosolvent can be involved in various *specific contacts* with DNA, which would disrupt the DNA hydration shell directly. In principle, cosolvent could interact both with the hydrophobic groups in the major groove (the thymine methyl groups and sugars), and with the hydrophilic groups N7, N6 and O4, thus shifting the equilibrium between the two hydration patterns shown in Figure 6b.

The opposite effects on the DNA curvature caused by the two osmolytes, glycerol and sucrose, under identical conditions (Pennings et al., 1992) indicate that the specific effects (II) may play at least some role. Clearly, further studies are necessary to elucidate the stereochemical mechanisms of the DNA-water-cations-cosolvent interactions and to resolve current discrepancies (compare different interpretations of the role played by MPD in decreasing the DNA curvature, suggested by Dlakic et al. (1996) and Dickerson et al. (1996)).

In conclusion, we wish to discuss the possible ways for experimental observation of the water tridents in the major groove of A-tracts predicted here.  First, the "trident water spine" in the major groove is expected to be more flexible than the Drew-Dickerson spine in the minor groove; the probability of the "trident" formation in the irregular $A_5T_5$ decamers is never more than 70%, and for the regular B' and $B^x$ forms it is only 25% in our simulations (Table 5).  This might be one of the reasons why the "trident" waters have not been detected in the AA:TT dimers so far.  Notice that the water trident W4' in the AG step of the *trp* operator (shown schematically in Figure 7e) is observed only when the *trp* repressor is bound to DNA -- in pure DNA crystal, this trident is not found (Shakked et al., 1994; Figure 3b).  By analogy, the water tridents predicted by us in the A-tracts might be more stable, and hence more easily detectable in the presence of the bound proteins (if the AA steps were open toward the major groove in such a case).

Second, the crystallographic observation of the water tridents bound to two consecutive base pairs in the major groove implies that in principle, locations of waters calculated by us are stereochemically feasible (compare Figures 7d-f).  We suggest that in solution the waters in the major groove of A-tracts could be localized directly using the $^{15}N$ NMR technique with [7-$^{15}N$]-labeled adenines (Gaffney et al., 1995).  Specifically, our calculations predict that at low temperature and in the absence of dehydrating agents the N7(Ade) "hydration index" (occupancy of the N7 atoms by waters) would be higher than at room or premelting temperatures and in the presence of alcohols.

Third, a new adenosine analogue, 7-hydroxymethyl-7-deazaadenosine was synthesized recently, and it structure was resolved by X-ray and NMR (Rockhill and Gumport, 1996).  Remarkably, the 7-hydroxy-group mimicking water bound to the N7 atom of adenine, was found in a position nearly identical to those calculated by us and shown in Figure 9, B'.  This opens interesting perspective of comparing the DNA bending and curvature under various conditions in natural A-tracts and in the synthetic tracts containing "pseudo-waters" covalently linked to the 7-th positions of purines.

Finally, the results of computations presented here are entirely consistent with the idea that the "structural" waters (Westhof, 1988) play an important role in determining the sequence-dependent conformational preferences of the DNA duplex.  These results, taken together with the crystallographic and solution data (Schneider and Berman, 1995; Brukner et al., 1994; Sprous et al., 1995) open a perspective for the comprehensive analysis of the sequence- and environment-dependent base-pair morphology (Gorin et al., 1995), in which waters and ions have to be considered in addition to the exocyclic groups of the bases, both in the minor and the major grooves.

## ACKNOWLEDGEMENTS

We thank R.L. Jernigan for general encouragement, support and critical reading of the manuscript. We are grateful to R.L. Dickerson, M. Dlakic, A.A. Gorin, R. Gumport , R.E. Harrington, S. Leikin, G.G. Malenkov, W.K. Olson, D.C. Rau, J. Rockhill and N.B. Ulyanov for valuable discussions and sharing with us their unpublished data. A.V.T. was partially supported by Fogarty International Center's Central and Eastern European Initiative. The authors acknowledge the National Cancer Institute for allocation of computing time and staff support at the Frederick Biomedical Supercomputing Center of the Frederick Cancer Research and Development Center.

**TABLE 1   Hydration energies for two regular $A_{10}$:$T_{10}$ duplex conformations, B' and B°**

| Conformation | B' | | B° | | Δ(B'-B°) | |
|---|---|---|---|---|---|---|
| Water content | 250 | 500 | 250 | 500 | 250 | 500 |
| $E_{total}$ | -227.3 | -447.7 | -223.6 | -443.7 | -3.7 | -4.0 |
| $E_W$ | -152.9 | -359.1 | -157.5 | -362.5 | 4.6 | 3.4 |
| $E_{W-DNA}$ | -74.4 | -88.6 | -66.1 | -81.2 | -8.3 | -7.4 |
| $E_{base}$ | -40.5 | -43.2 | -36.9 | -38.4 | -3.6 | -4.8 |
| $E_{sug}$ | -9.8 | -12.9 | -9.0 | -12.5 | -0.8 | -0.4 |
| $E_{phos}$ | -24.1 | -32.5 | -20.2 | -30.3 | -3.9 | -2.2 |

Δ(B'-B°) refers to the energy difference between the two conformations, B' (obtained with the NMR restraints), and B° (with no restraints) at T=300 K, see Methods. Δ(B'-B°) = 0.9 kcal/mol for the conformational energy of DNA *in vacuo*  (Zhurkin et al., 1990).

$E_W$ is the energy of water-water interactions;  $E_{W-DNA}$ is for the DNA-water interactions; $E_{total}$ is $E_W$+$E_{W-DNA}$. $E_{base}$ , $E_{sug}$ , $E_{phos}$  denote the energy of interactions between the waters and bases, sugars, phosphates, respectively.  In this and in the following tables the energy values are given in kcal/mol base pairs.

**TABLE 2** Hydration energies of bases and hydration indexes of hydrophilic atoms for two $A_{10}:T_{10}$ duplexes

| Conformation | B' | B° | Δ(B'-B°) |
|---|---|---|---|
| *Hydration energy* | | | |
| E(A) | -24.9 (0.5) | -21.5 (0.7) | -3.4 |
| E(T) | -18.3 (0.4) | -16.9 (0.4) | -1.4 |
| $E_m$(A) | -7.9 (0.2) | -8.0 (0.4) | 0.1 |
| $E_m$(T) | -5.7 (0.2) | -4.8 (0.3) | -0.9 |
| $E_M$(A) | -15.5 (0.4) | -12.1 (0.5) | -3.4 |
| $E_M$(T) | -10.2 (0.3) | -9.4 (0.3) | -0.8 |
| *Hydration index* | | | |
| N3 | 0.97 (0.01) | 0.96 (0.08) | 0.01 |
| O2 | 0.97 (0.02) | 0.85 (0.04) | 0.12 |
| N7 | 1.66 (0.08) | 1.30 (0.08) | 0.36 |
| H(N6) | 0.84 (0.04) | 0.78 (0.06) | 0.06 |
| O4 | 1.15 (0.06) | 0.96 (0.04) | 0.19 |

Conditions: 500 water molecules at 300 K; r.m.s. deviations are given in parentheses.

E(A) and E(T), total adenine and thymine hydration energies (kcal/mol of bases);

$E_M$(A) and $E_M$(T), the major groove hydration energies;

$E_m$(A) and $E_m$(T), the minor groove hydration energies for adenine and thymine.

Only those water molecules were taken into account which were located not farther than 7.5 Å from the center of a base, therefore $E_M(A)+E_m(A) \neq E(A)$.

**TABLE 3   Hydration energies for two A5T5 duplex conformations, B' and B°**

| Conformation | B' | B° | Δ(B'-B°) |
|---|---|---|---|
| $E_{total}$ | -226.8 | -222.7 | -4.1 |
| $E_W$ | -153.5 | -157.1 | 3.6 |
| $E_{W-DNA}$ | -73.4 | -65.6 | -7.8 |
| $E_{base}$ | -40.4 | -37.6 | -2.8 |
| $E_{sug}$ | -9.8 | -8.4 | -1.4 |
| $E_{phos}$ | -23.1 | -19.6 | -3.5 |
| $E(A)$ | -23.3 | -21.1 | -2.2 |
| $E(T)$ | -17.1 | -16.5 | -0.6 |
| $E_m(A)$ | -7.3 | -7.4 | 0.1 |
| $E_m(T)$ | -5.5 | -4.7 | -0.8 |
| $E_M(A)$ | -14.8 | -12.3 | -2.5 |
| $E_M(T)$ | -9.7 | -9.4 | -0.3 |

Conditions:  250 water molecules at 300 K.  Notations as in Tables 1 and 2.

**TABLE 4   Structural features of the water shell in the minor groove of $A_{10}:T_{10}$ duplexes**

| Conformation | B' | | $B^x$ | | B° | |
|---|---|---|---|---|---|---|
| Minor groove width (Å) | 6.4 | | 7.2 | | 8.7 | |
| Water bridge | $D$ (Å) | $P$(%) | $D$ (Å) | $P$(%) | $D$ (Å) | $P$(%) |
| N3···O2  (*m1*) | 3.6 | 92 (2) | 4.0 | 76 (6) | 4.2 | 21 (7) |
| Spine (two layers) | - | 22 (13) | - | 15 (6) | - | 0 |
| N3···O2 two-water  (*m2*) | 3.6 | 0 | 4.0 | 3 (2) | 4.2 | 31 (5) |
| O4'···N3  (*m3*) | 3.7 | 0 | 3.8 | 0 | 4.5 | 21 (7) |
| O4'···O2  (*m3*) | 3.3 | 0 | 3.5 | 0 | 3.6 | 7 (2) |
| N3···N3 | 4.2 | 0 | 4.0 | 0 | 4.2 | 0 |
| O2···O2 | 4.3 | 0 | 4.1 | 0 | 4.2 | 0 |

Conditions:  500 water molecules at 300 K.  $D$ is the distance between the two hydrophilic atoms, $P$ is the probability of occurrence of the corresponding hydration pattern;  r.m.s. deviations are given in parentheses.  $B^x$ is the average X-ray structure of the A-tract (Figure 1). The groove width is measured as the O4'--O4' distance (Figure 3).  The water bridges *m1, m2, m3* are defined in Figure 3.  Hydration spine is defined as a simultaneous formation of six hydrogen bonds between three waters and four atoms N3/O2: two bridges of the *m1* type, linked by the second layer water (Drew & Dickerson, 1981).  Only those occurrences were included in the printout, whose probabilities were 3% and higher; thus zero means that the actual value of $P$ is less than 3%.

**TABLE 5** Structural features of the water shell in the major groove of $A_{10}$:$T_{10}$ duplexes

| Conformation | | B' | | $B^x$ | | B° | |
|---|---|---|---|---|---|---|---|
| Major groove width (Å) | | 17.1 | | 15.0 | | 14.5 | |
| Water pattern | | $D$ (Å) | $P$ (%) | $D$ (Å) | $P$ (%) | $D$ (Å) | $P$ (%) |
| N7⋯H(N6) | *(M1)* | 4.7 | 46 (9) | 4.1 | 61 (10) | 3.8 | 42 (10) |
| *N7⋯H(N6) | *(M2)* | 3.1 | 38 (7) | 3.1 | 30 (11) | 3.1 | 21 (7) |
| N7⋯N7 | *(M3)* | 4.3 | 32 (12) | 4.1 | 24 (9) | 4.2 | 9 (4) |
| Tridents | | - | 26 (9) | - | 24 (9) | - | 9 (4) |
| O4⋯H(N6) | *(M4)* | 4.2 | 0 | 3.3 | 15 (7) | 3.5 | 19 (6) |
| O4⋯O4 | *(M5)* | 3.9 | 13 (4) | 3.7 | 4 (2) | 3.7 | 3 (2) |

Conditions: 500 water molecules at 300 K. $D$ is the distance between the two hydrophilic atoms, $P$ is the probability of occurrence of the corresponding hydration pattern; r.m.s. deviations are given in parentheses. The groove width is measured as the C2'⋯C2' distance (Figure 7). The water bridges *M1-M5* and the tridents are defined in Figure 7.

\* In the case of *M2* bridge both atoms N7 and H(N6) belong to the same adenine.

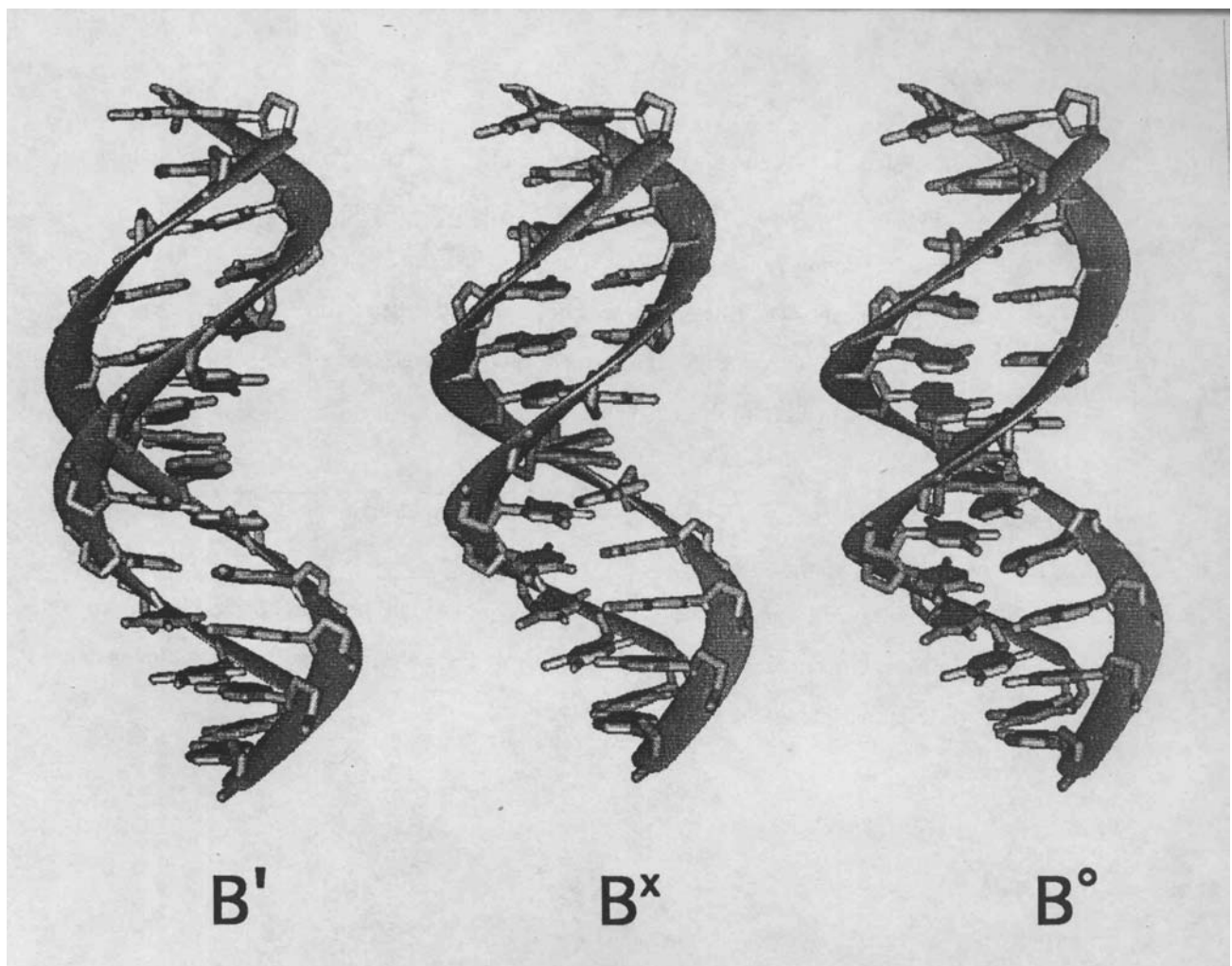

Figure 1. Three $A_{10}$:$T_{10}$ decamer conformations used in the calculations: B' form consistent with the NMR restraints at low temperature (Nadeau & Crothers, 1989); $B^x$ form representing the average conformation of the AA:TT steps in the crystal B-DNA structures (Gorin et al., 1995); B° form averaged over the MC ensemble generated without the NMR restraints (Zhurkin et al., 1990). Notice that the minor groove width increases from left to right as the roll angle increases from -6.4° to 0.6° to 4.6° respectively. The roll angle definition is given in (Gorin et al., 1995); this is consistent with the EMBO Convention (Dickerson et al., 1989), so the base-pairs are rolled toward the minor groove in B', and toward the major groove in B°. The inter-proton distances characterizing the minor groove size, (Ade)H2⋯H1'(Thy) and (Ade)H2⋯H1'(Ade) (given in parentheses), are as follows: B' form, 3.70Å (4.01); $B^x$ form, 4.14Å (4.84); B° form, 4.07Å (5.45). To demonstrate the periodic boundary conditions imposed on the system during MC simulation, 12 base pairs are shown. Figures 1, 6, 9 were generated using MIDAS (Ferrin et al., 1988; Huang et al., 1991).

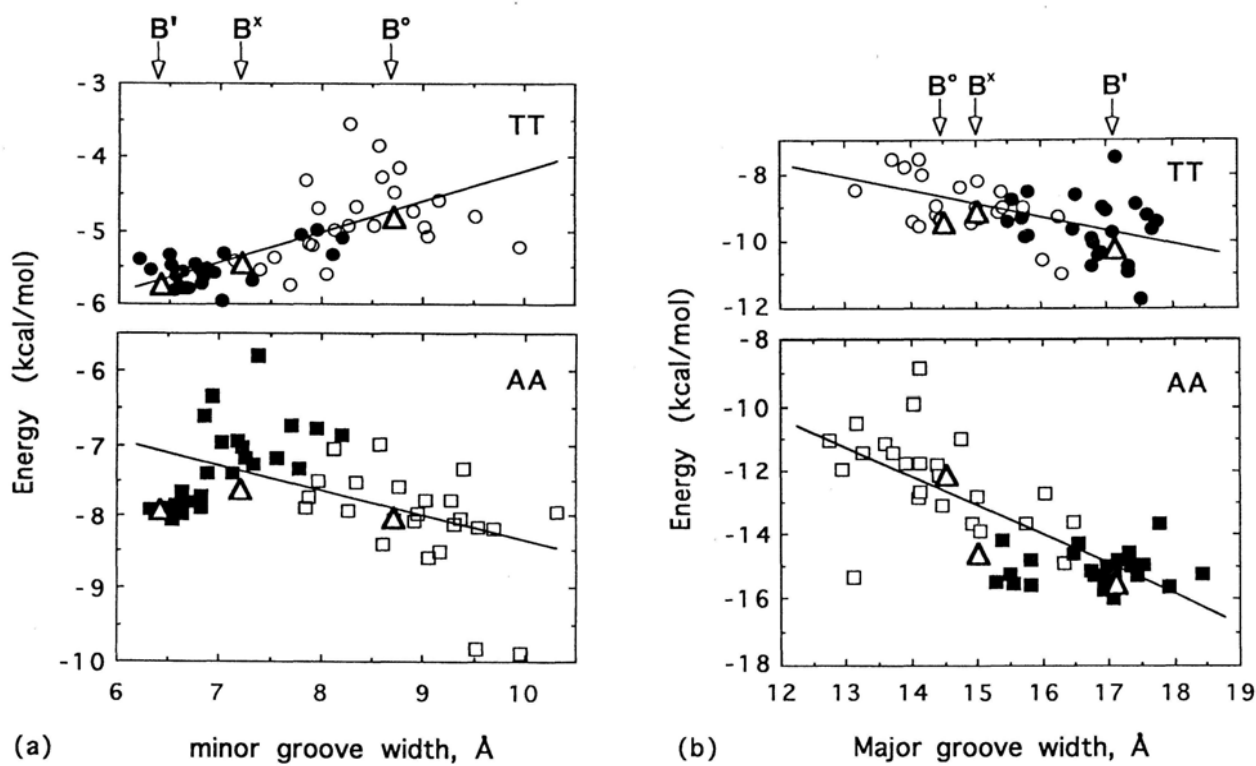

Figure 2.

(a) The dependence of the minor groove hydration energy on the minor groove width for the eight $A_5T_5$ decamers (see Methods).

(b) The dependence of the major groove hydration energy on the major groove width.

The groove widths are measured as the distances O4'···O4' (a) and C2'···C2' (b) as shown in Figures 3 and 7 respectively. The hydration energies are calculated separately for the thymines (circles) and for the adenines (squares). The data for $A_5T_5$ decamers generated with the NMR restraints are shown as filled symbols (narrow minor groove and wide major groove), and those without NMR restraints are shown as open symbols. To make the data comparable with the results for $A_{10}$:$T_{10}$ decamers, in each run of $A_5$ or $T_5$ only the three central bases were taken into account. The data for the three $A_{10}$:$T_{10}$ decamers (B', $B^x$ and B°) are shown as triangles (see Tables 2, 4 and 5). The best-fit lines correspond to the data for $A_5T_5$ only.

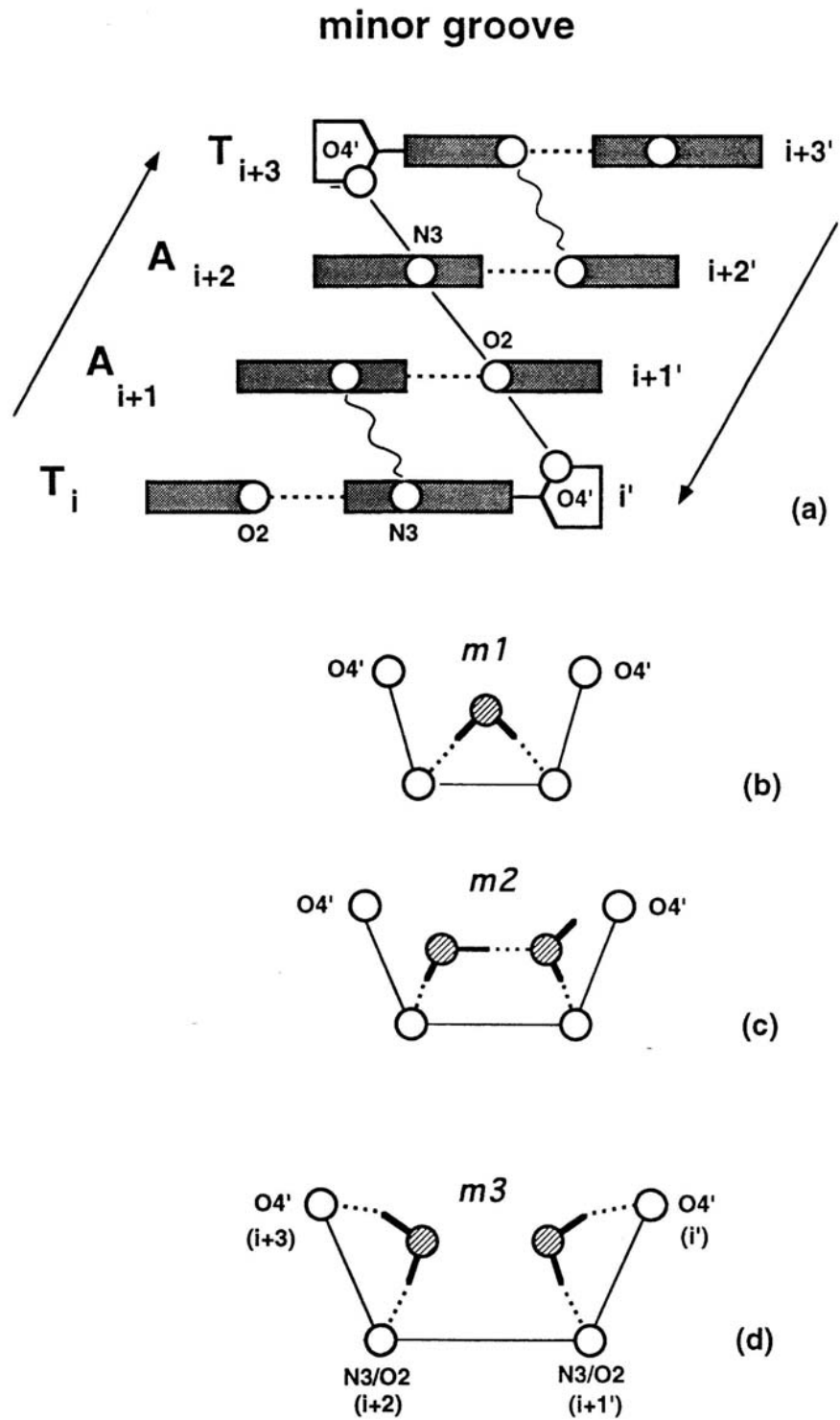

Figure 3. Minor groove scheme (a) and hydration patterns *m1*, *m2* and *m3* (b, c, d). The hydrogen bond acceptors are shown as open circles. The groove profiles in (b, c, d) represent crossections through the atoms O4'(i+3)-N3(i+2)-O2(i+1')-O4'(i') (a). Notice that as the groove becomes wider in the order (b, c, d), the preferable hydration pattern changes correspondingly from *m1* to *m2* to *m3*. Sequence TAAT is chosen to illustrate that all the possible AT-containing steps have a similar arrangement of the hydrogen bond acceptors: the profiles (b, c, d) are roughly the same for the TA, AA and AT steps. The sequence chain is numbered in the 5'-3' direction according to the EMBO Convention (Dickerson et al., 1989).

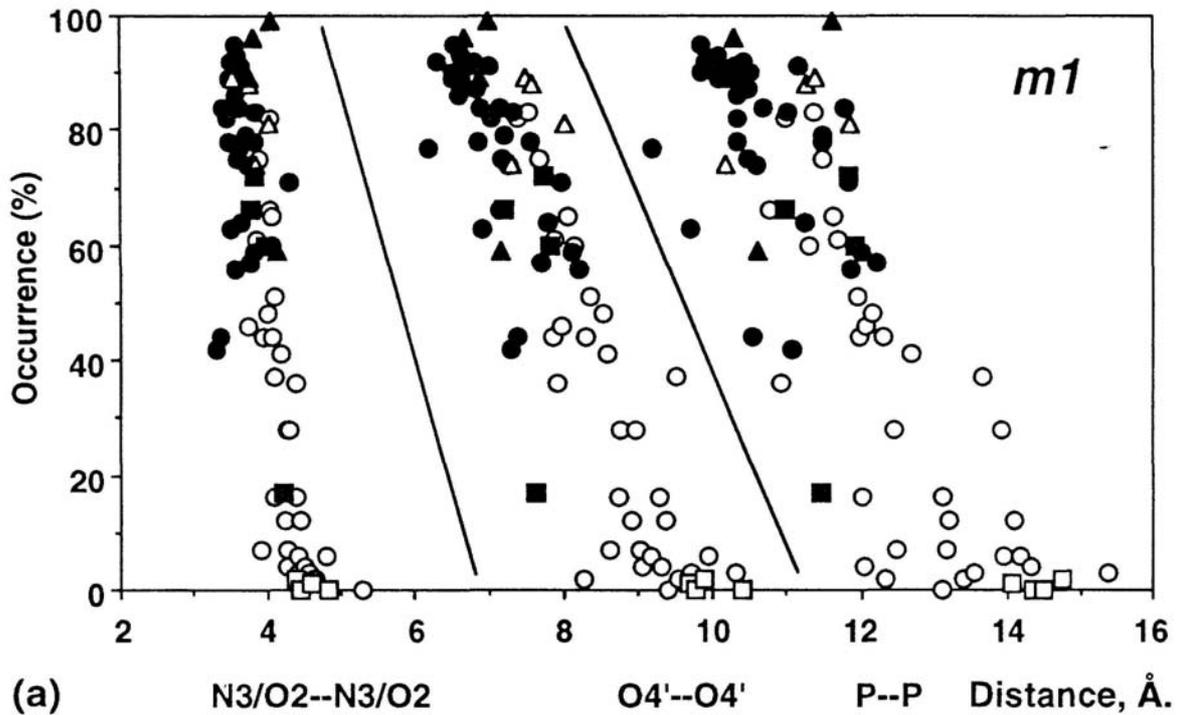

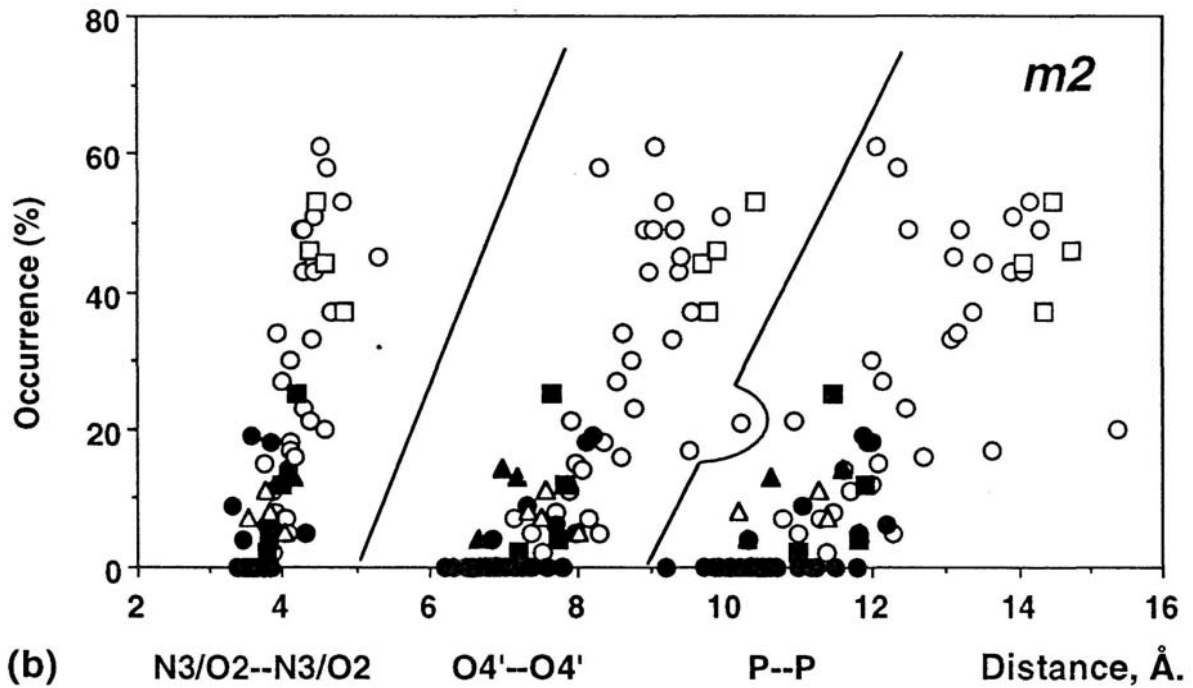

Figure 4. Probabilities of the base-base one-water bridge *m1* (a) and two-water bridge *m2* (b) *versus* the minor groove width for the $A_5T_5$ decamers. The groove width is evaluated in three ways: as the distance between the two N3/O2 atoms from the adjacent bases (i+1') and (i+2); as the distance between the two deoxyribose oxygens, O4'(i+3) and O4'(i'), see Figure 3a; as the distance between the two closest phosphate atoms from the opposite strands, P(i+3) and P(i-1'). The data for the AA dimeric steps are shown by circles, for AT by triangles and for TA by squares. The data for decamers generated with the NMR restraints (B' form) are shown by filled symbols, and without NMR restraints (B° form) by open symbols.

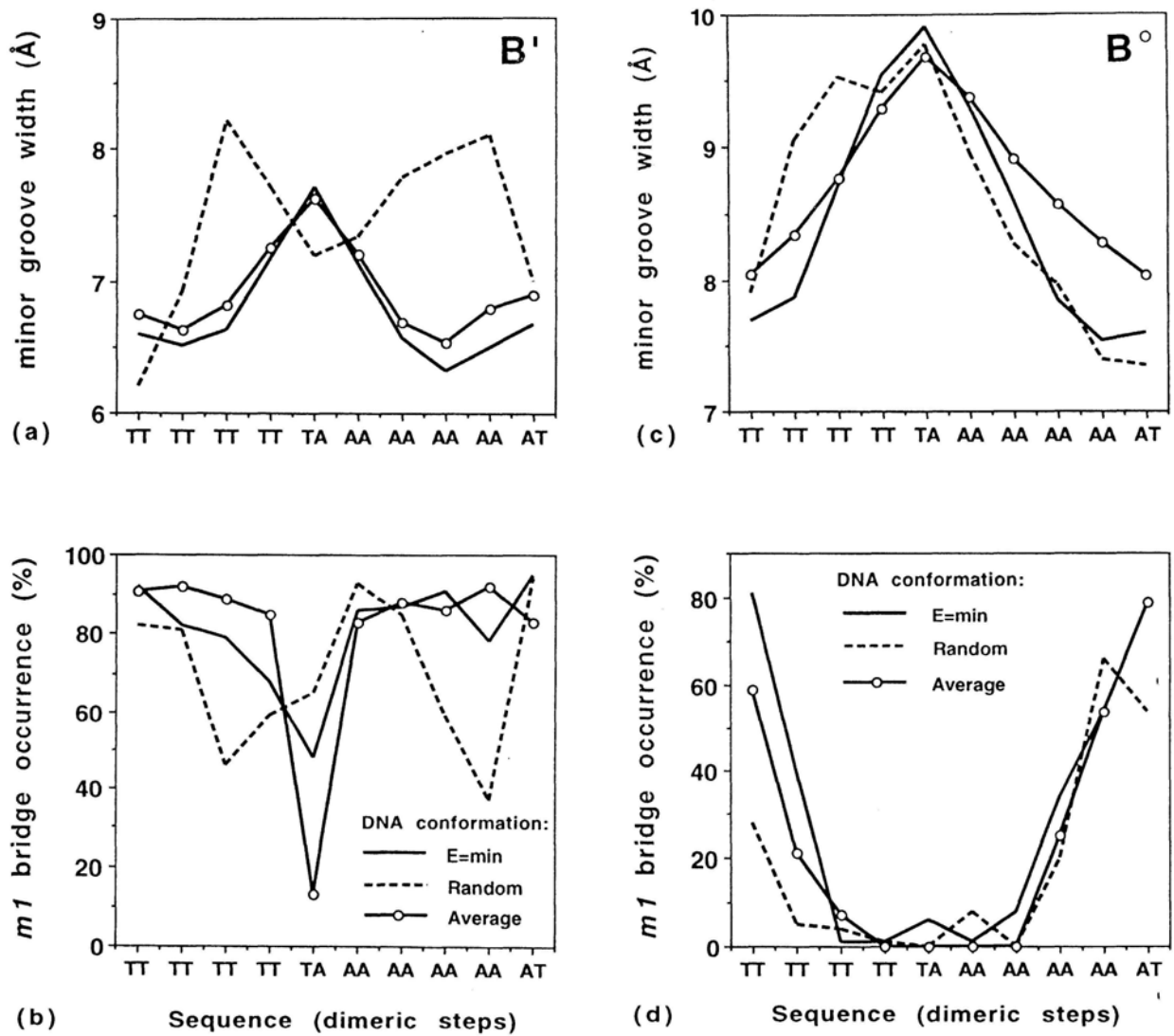

Figure 5. Profiles of the minor groove widths and the water bridge *m1* probabilities as functions of sequence for $A_5T_5$ decamers generated with the NMR restraints, B' form (a, b) and without restraints, B° form (c, d). The data for the energy minimum and the averaged conformations of $A_5T_5$ are shown by solid lines and by the lines with circles respectively; the "instantaneous" conformations are represented by the broken lines. The minor groove width is measured as the distance between the two closest deoxyribose oxygens O4' across the groove, see Figure 3a.

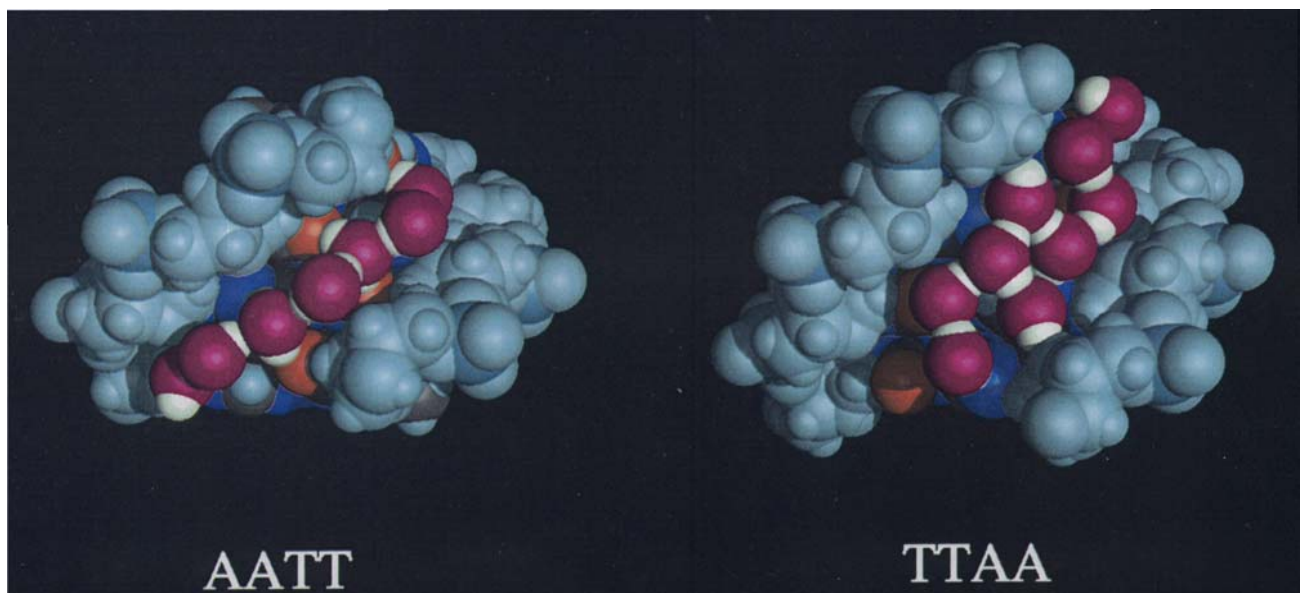

Figure 6. (a) Hydration patterns in the minor groove: hydration spine in the AATT tetramer and water strings in TTAA.

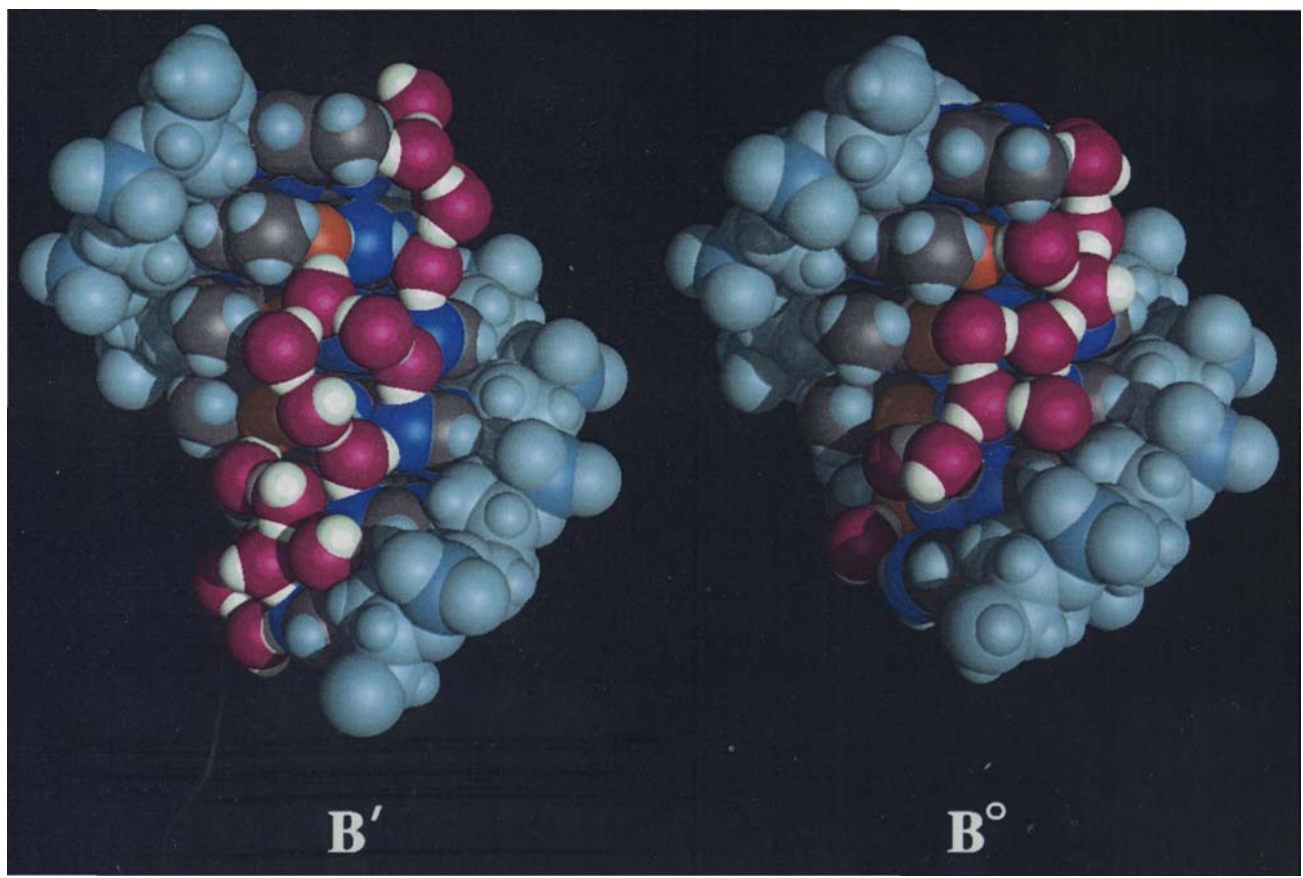

Figure 6. (b) Hydration patterns for the "wide" (B') and the "normal" (B°) major grooves of $A_{10}:T_{10}$.  The waters are shown in magenta and white; the nitrogens and oxygens of bases are in blue and orange respectively; the sugar-phosphate backbone is in gray.  In (b), the aromatic carbons and C5(Met) of thymines are shown in dark gray.  Notice that the methyl groups, together with the H(C6)(Thy) and $CH_2$ groups of sugars, form a prolonged hydrophobic cluster.

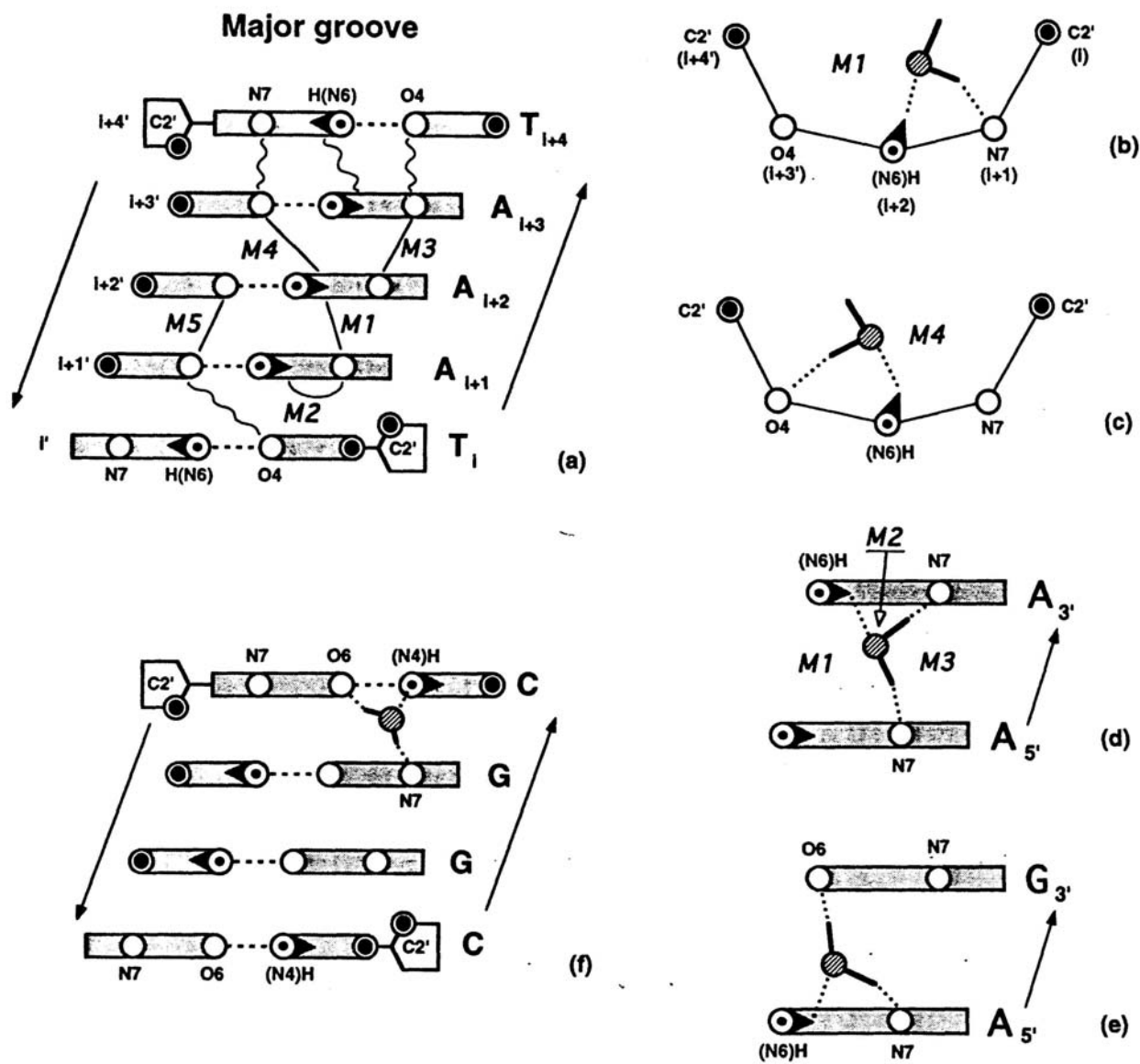

Figure 7. Major groove scheme (a), water bridges *M1* (b), *M4* (c) and water trident (d). Each of the bridges *M1-M5* is shown in (a) only once. The most probable water bridges occurring in the AT and TA steps are shown by wavy lines (a). The profiles of the groove in (b, c) show the crossection through C2'(i+4')-*M4-M1*-C2'(i) in (a). The water molecule linking together two adenines is denoted as a trident when it forms three bridges *M1*, *M2* and *M3* simultaneously (d).
(e) Water trident observed in the AG step in the *trp* repressor/operator complex (water W4' in Figure 3b, Shakked et al., 1994). Notice different orientations of the waters in (d) and (e).
(f) Major groove scheme for GC-containing sequence is shown for comparison, together with the water trident observed in the GC step in B-DNA dodecamer (water #38 in Figure 10 and Table 2, Drew and Dickerson, 1981). The hydrogen bond acceptors O4, O6 and N7 are shown as open circles, the hydrophobic groups $H_2$(C2'), Met(Thy) and H(C5)(Cyt) as filled circles, and the donors N4(Cyt) and N6(Ade) as circles with dots. The amino-protons H(N4) and H(N6) not involved in the Watson-Crick hydrogen bonds and directed toward the groove are shown as black triangles.

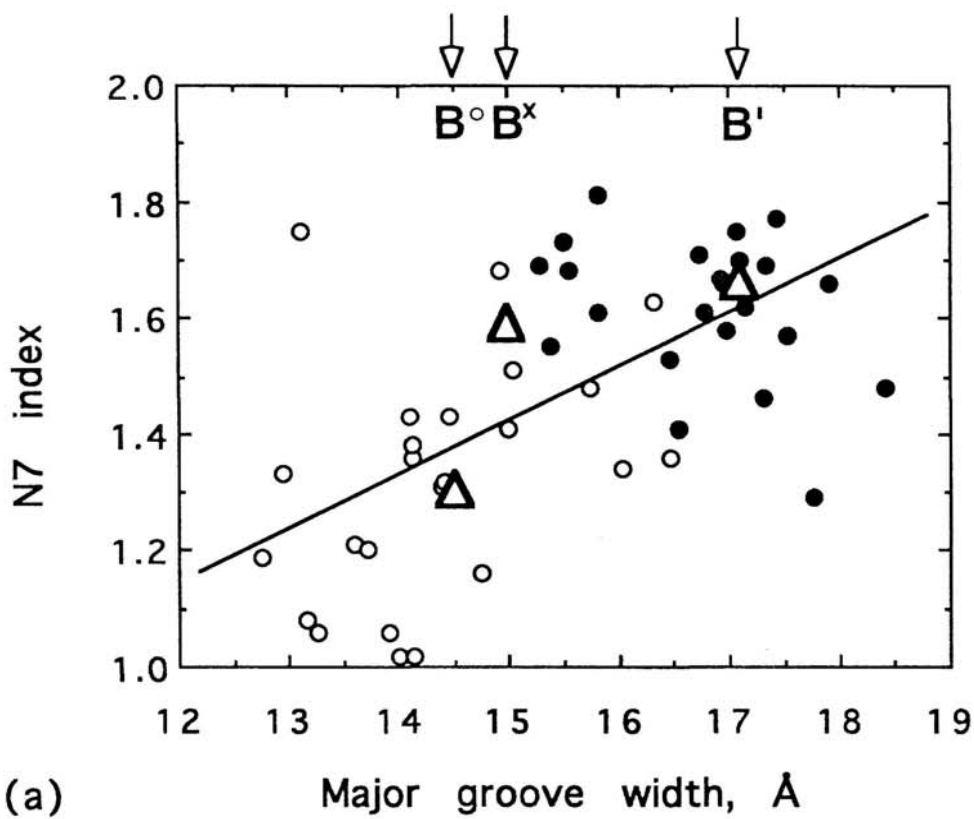

(a)

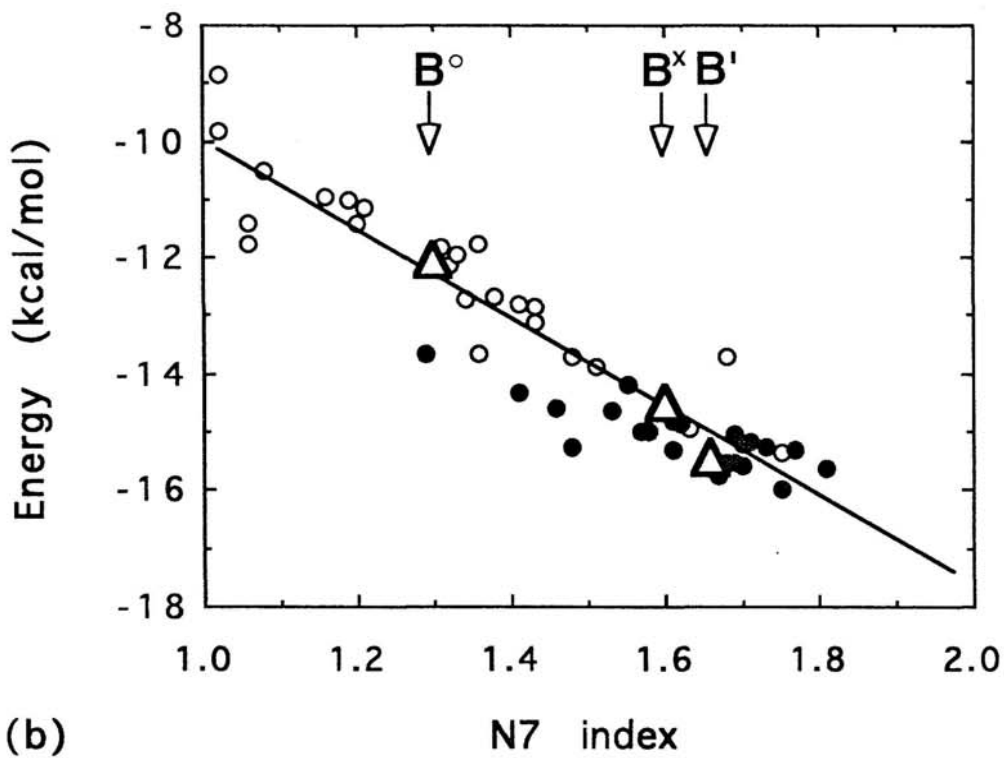

(b)

Figure 8. Correlations between the major groove width, the N7(Ade) hydration index and the hydration energy of adenines in the major groove. The groove width is measured as the C2'···C2' distance (Figure 7a). As in Figure 2, only the three middle adenines in the runs of $A_5$ were taken into account. The filled and open circles show the data for the $A_5T_5$ decamers obtained with and without NMR restraints respectively. Triangles refer to $A_{10}$:$T_{10}$ conformations B°, B$^x$ and B' (Tables 2, 4 and 5).

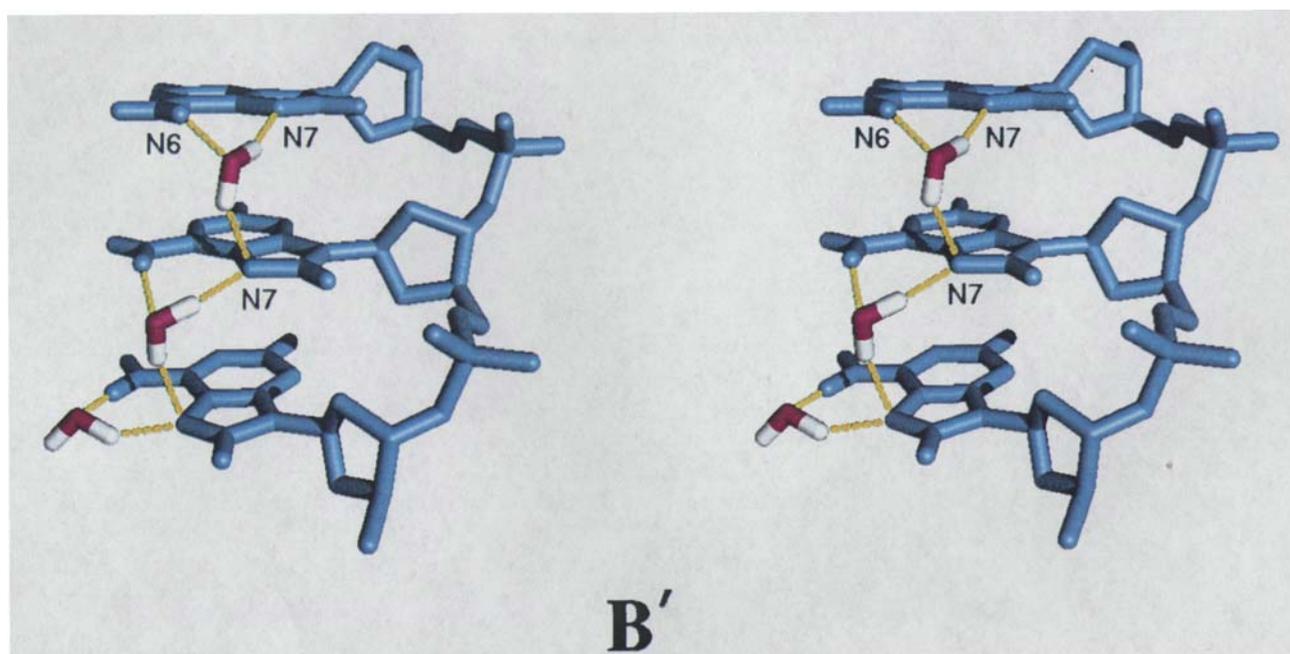

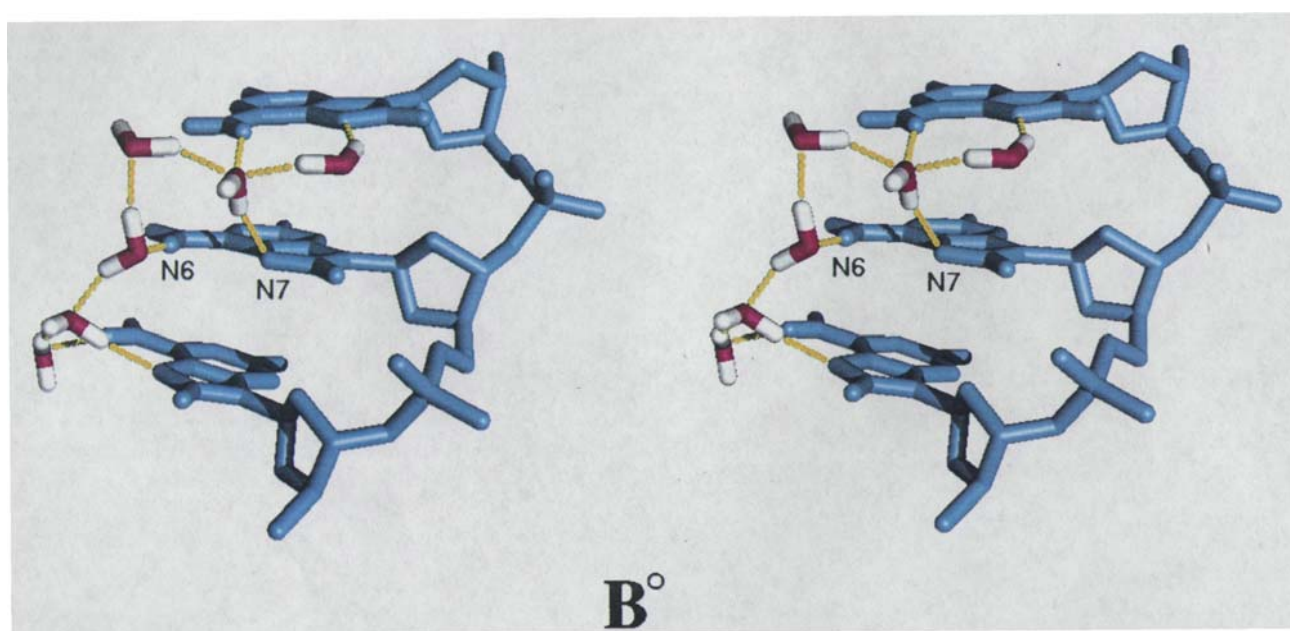

Figure 9. Stereo representations of the hydration patterns in the major groove shown in Figure 6b. The waters are in magenta and white; DNA is in cyan; only adenines are shown. Notice pseudo-regular positioning of the water tridents in the "wide" groove in B' form as opposed to less structured water chains in the "normal" major groove in B° form. Each water trident in B' is hydrogen bonded to two N7 acceptors and to one H(N6) donor.